# Influence of the magnetic field's curvature on the radial-azimuthal dynamics of a Hall thruster plasma discharge with different propellants

M. Reza*[1], F. Faraji*, A. Knoll*

*Plasma Propulsion Laboratory, Department of Aeronautics, Imperial College London, London, United Kingdom

**Abstract**: The topology of the applied magnetic field is an important design aspect of Hall thrusters, which affects the plasma behavior and the propulsive performance of the device. For modern Hall thrusters, the magnetic field topology most often features curved lines with a concave (negative) curvature upstream of the magnetic field peak and a convex (positive) curvature downstream. Additionally, the advent of the magnetic shielding technique has resulted in the design of Hall thrusters with non-conventional magnetic field topologies that exhibit high degrees of concavity upstream of the field's peak. In this article, we carry out a rigorous and detailed study of the effects that the magnetic field's curvature has on the plasma properties and the underlying processes in a 2D configuration representative of a Hall thruster's radial-azimuthal cross-section. The analyses are performed for plasma discharges of three propellants of high applied interest: xenon, krypton, and argon. For each propellant, we have carried out high-fidelity reduced-order quasi-2D particle-in-cell (PIC) simulations with various degrees of positive and negative curvatures of the magnetic field. Corresponding 1D radial PIC simulations were also performed for xenon to compare the observations between 1D and 2D simulations. Most notably, we observed that there are distinct differences in the plasma phenomena between the cases with positive and negative field curvatures. The instability spectra in the cases of positive curvature is mostly dominated by the Electron Cyclotron Drift Instability, whereas the Modified Two Stream Instability is dominant in the negative-curvature cases. The distribution of the plasma properties, particularly the electron and ion temperatures, and the contribution of various mechanisms to electrons' cross-field transport showed notable variations with the field's curvature, especially between the positive and the negative values. Finally, the magnetic field curvature was observed to majorly influence the ion beam divergence along the radial and azimuthal coordinates.

**Section 1: Introduction**

Hall thrusters are a plasma propulsion technology that have an extensive heritage of in-space application and are today prime candidates to access the next generation of near-Earth and interplanetary missions. The great applied interest in the Hall thruster technology has been because, on the one hand, these thrusters provide specific impulse values of at least an order of magnitude higher than the chemical thrusters, thus, providing huge mass advantages at the satellite level. On the other hand, Hall thrusters feature a unique combination of high efficiency, high thrust-to-power ratio, and operational versatility compared to other electric propulsion technologies on the market.

The plasma in a Hall thruster is subject to a perpendicular configuration of the electric ($E$) and magnetic ($B$) fields, a configuration which is referred to as cross-field or E×B. The electric field is self-consistently generated and maintained, whereas the magnetic field is externally applied. The intensity of the magnetic field is such that the electrons are fully magnetized, but the ions are mostly non-magnetized and are accelerated under the influence of the electrostatic field. The topology and the magnitude of the $B$-field plays a significant role in the performance and the operational physics of the thrusters. As the dynamics of the magnetized electrons is influenced by the $B$-field, the processes and phenomena which are driven and/or affected by the properties and the behavior of the electron species, from the ionization and acceleration to the plasma-wall interactions and instabilities, can be impacted by the applied magnetic field. This emphasizes the critical importance of a proper design of the magnetic field topology in terms of its spatial distribution and magnitudes to achieve high-performance and stable operation of Hall thrusters.

In modern Hall thrusters, the magnetic field features a strong gradient along the axial direction of the thruster's discharge channel, parallel to its centerline, and a moderate radial gradient along the radius of the channel [1][2]. The field lines have a concave curvature along the axial-radial plane upstream of the axial location of the magnetic field peak intensity, which typically occurs near the exit plane of the channel. Downstream of the magnetic field peak, the field lines have a convex curvature. About the location of the $B$-field peak, the field lines are mostly radial with no notable curvature. The concave (negative) curvature of the field lines is often referred to as being "ion focusing" in the literature [2]. On the contrary, the field lines with convex (positive) curvatures are known as "ion diverging" [2]. In recent years, the technique of magnetic shielding [3][4] of the channel walls has been applied across the Hall thrusters of various power levels [5]-[9] in order to mitigate and/or eliminate the life-limiting issue of the channel erosion [3]. The so-called "magnetically shielded" thrusters feature $B$-field topologies with large degrees of negative curvature upstream of the magnetic field peak and, hence, strong radial gradients [2][3].

---

[1] **Corresponding Author** (m.reza20@imperial.ac.uk)

The effects of the axial gradients of the magnetic field on the plasma behavior in Hall thrusters have been the subject of several studies in the past years [10]-[14]. However, there has not been as much attention to the effects that the magnetic field curvature (radial gradients) have on the underlying plasma processes, particularly on the instabilities and the electron transport. Most high-fidelity studies on these two phenomena, which involved the radial coordinate, had been carried out in simplified configurations that assume perfectly radial magnetic field lines [15]-[17].

Concerning the impacts of the magnetic field curvature (and the associated radial gradients) on the plasma-wall interactions, there have been, nonetheless, a number of previous efforts. These activities were in part motivated by the emergence of the magnetically shielded Hall thrusters and their non-conventional magnetic field topologies. Refs. [3][18][19] carried out hybrid fluid-PIC simulations in the axial-radial coordinates to investigate the physics underlying the magnetic shielding concept as well as to analyze the variation in the distribution of the plasma properties, the sheath characteristics, and the plasma-wall interactions due to the consequent changes in the $B$-field topology. Refs. [20][21] also performed studies of similar scope on two specific industrial magnetically shielded Hall thrusters using axial-radial hybrid simulations. The focus of these works were on the 2D spatial distribution of the plasma properties, global power and current balances, and the interactions of the plasma with the radial walls [20][21].

The above effects of the magnetic field curvature has also been assessed using fully kinetic PIC simulations. In this regard, Refs. [22][23] studied the influence of various degrees of field's curvature on the plasma-wall interactions and the erosion phenomenon using 2D axial-radial and limited-domain radial-azimuthal simulations, respectively. They demonstrated that increasing the concavity of the field lines reduces the energy of the ions reaching the wall and, hence, lowers the erosion rate [23]. On the contrary, in the case of convex field lines, the erosion rate becomes highly significant at large degrees of positive curvature [23]. In Ref. [24], the authors carried out 1D radial PIC simulations with various magnetic field curvatures and showed that, for negative curvatures, the electrons' distribution function becomes more isotropic, and the tail of the distribution function is greatly replenished. A concentrating effect in cases with concave field curvatures was observed, with the plasma density in the center of the domain becoming higher than the case with a constant radial magnetic field [24]. Ref. [25] assessed, again in a 1D radial setup, the effects of the magnetic field's inclination on the plasma sheath's characteristics in the presence of the secondary electron emission and with the plasma properties representative of different axial locations within the discharge channel of a Hall thruster.

In a previous publication of direct relevance to the present work [26], we performed 2D radial-azimuthal PIC simulations of an E×B discharge of the xenon gas for two specific $B$-field configurations representative of the magnetic shielding topology. We focused, in that effort, on the variation in the instability spectra between these two configurations and the resulting changes in the dominant mechanism behind the electrons' axial mobility [26]. We most notably showed that, in the presence of a strong magnetic field gradient, a long-wavelength wave mode develops due to an inverse energy cascade and induces a significant electron cross-field transport [26].

Despite the important insights derived, the setup of the simulations in Ref. [26] was such that the $B$-field curvature could not be systematically varied in order to fully quantify the correlations between the degree of the field's curvature and the change in the plasma processes. Moreover, both field configurations featured negative concave curvatures [26]. Accordingly, inspired by the methodology pursued in Refs. [23][24], and building upon our prior work [26], we present here in a 2D radial-azimuthal setup a detailed and rigorous study into the effects of negative and positive magnetic field curvatures on the discharge's behavior and dynamics. In the present research, we have additionally extended the investigations of Ref. [26] to three different propellants, namely, xenon, krypton, and argon. For various propellants and degrees of $B$-field's curvature, we discuss the variations induced in the spectra of the plasma instabilities and, consequently, in the distribution of the plasma properties, the electron transport characteristics, and the divergence of the ion beam.

To the best of our knowledge, the parametric numerical studies of this work and the insights derived as a result have not been reported in the existing literature. We were further motivated to publish our findings in two additional respects: (1) the important differences that we noticed concerning the effects of the field's curvature between the 1D and the 2D studies, which highlights the critical significance of the azimuthal physics, (2) the distinct changes that we observed, when comparing the cases of negative and positive curvature, in the characteristics of the dominant instabilities, which is a novel observation that can have significant implications in the basic understanding of the plasma processes within Hall thrusters.

**Section 2: Simulations' setup and conditions**

We performed a set of quasi-2D (Q2D) simulations using the reduced-order PIC code of the Imperial Plasma Propulsion Laboratory (IPPL), which is named IPPL-Q2D [27][28]. The domain decomposition associated with the reduced-order PIC scheme [27][28] was carried out using 50 regions along each of the involved simulation dimensions. More details on the verification results of the 50-region quasi-2D simulation, its computational advantage, and the physics studies enabled via this specific approximation order of the reduced-order PIC can be found in Refs. [26][28][29].

The domain of the Q2D simulations represent a radial-azimuthal cross-section of a Hall thruster. Corresponding simulations are performed in a 1D setting along the radial direction of the thruster's domain as well using the IPPL-1D code [30][31] to enable the comparison of the results from the Q2D simulations. Such comparison allows distinguishing the observations that are mainly driven by the azimuthal dynamics from the processes along the radial direction directly associated with the plasma-wall interactions. In the following, we introduce the setup and the conditions of the Q2D simulations. The same setup is adopted for the 1D simulations except for the fact that the azimuthal direction and all associated conditions are excluded. This has been necessary to make sure that a direct comparison between the results from the Q2D and the 1D simulations is possible.

The setup of the simulations is, overall, similar to the one corresponding to the radial-azimuthal benchmark case of Ref. [15]. The domain represents a Cartesian plane whose extents along the radial direction ($L_x$) and the azimuthal direction ($L_z$) are equal to 1.28 cm. In these simulations, $x$, $y$, and $z$ denote, respectively, the radial, the axial, and the azimuthal coordinates. A constant and spatially uniform axial electric field ($E_x$) with the intensity of 10 $kVm^{-1}$, and a constant radial magnetic field ($B_x$) with the magnitude of 20 mT are imposed.

To investigate the impact of the curvature of the magnetic field on the dynamics of the plasma in the adopted simulation configuration, the total applied magnetic field comprises a radially varying axial component ($B_y$) in addition to the radial component. We chose the same relation adopted in Ref. [24] to prescribe the variation of the $B_y$ along the radial direction

$$B_y = B_x \tan\theta(r); \quad \tan\theta = \frac{L_x - x}{L_x}\tan\theta_1 + \frac{x}{L_x}\tan\theta_2, \quad \text{(Eq. 1)}$$

where, $\theta_1$ and $\theta_2$ are, respectively, the incident angles of the magnetic field lines at the inner and the outer walls, as illustrated in Figure 1. The magnetic field is assumed to be symmetric with respect to the channel centerline, hence, $\theta_1 = -\theta_2$. We highlight that, because the simulations occur on a plane and do not resolve the variations along the axial dimension, the magnetic field does not have an axial gradient.

The Q2D simulations are performed with three different propellants, namely, xenon, krypton, and argon, and with various degrees of magnetic field curvature, as parametrized by $\theta = \theta_1$. The field curvature parameter, $\theta$, ranges from -60 to 60 degrees with the increment of 15 degrees. The case with $\theta = 0$ corresponds to the purely radial magnetic field topology of the benchmark setup. The counterpart 1D simulations are conducted only for xenon.

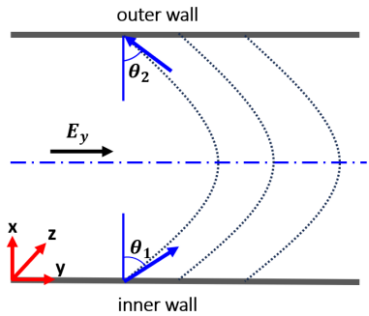

Figure 1: Schematic of the magnetic field lines topology in the axial-radial ($x - y$) plane, with the lines' incidence angle at the inner and outer wall, $\theta_1$ and $\theta_2$, used in Eq. 1 illustrated.

The negative curvatures refer to the field lines curving toward the anode (-$y$ direction), which is representative of a typical magnetic field topology upstream of the axial location of the magnetic field peak ($x_{B,peak}$) within a Hall thruster's discharge channel. Conversely, the positive curvatures signify the field lines curving toward the cathode (+$y$ direction), hence, resembling the conditions downstream of the $x_{B,peak}$ location. The radial variation of $B_y$ and the streamlines of the magnetic field in the simulated cases are illustrated, respectively, in Figure 2(a) and (b).

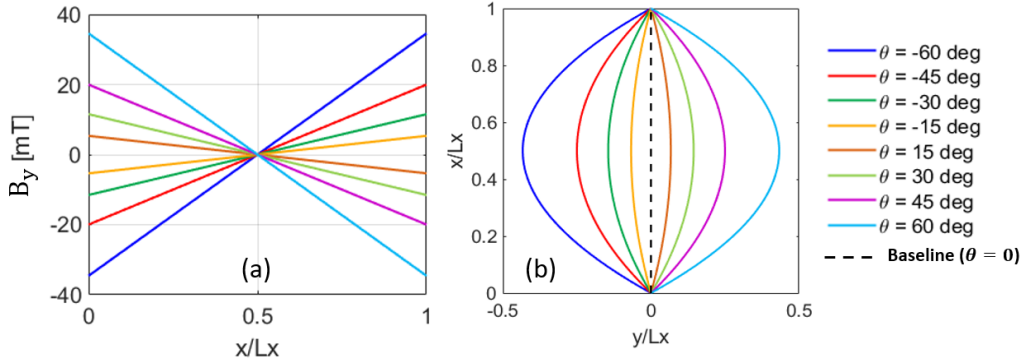

Figure 2: (a) Radial profile of the axial component of the magnetic field ($B_y$) for various degrees of field curvature ($\theta$). (b) 2D shape of the magnetic field lines in the axial-radial ($y$-$x$) plane for various values of $\theta$.

Regarding the numerical parameters, for the simulations with positive field curvatures and negative field curvatures of -30 and -15 degrees, the cell size (cell number) is 50 $\mu m$ (256 nodes) along both the radial and azimuthal directions, and the time step is $1.5 \times 10^{-11}$ s.

| | Value [unit] | |
|---|---|---|
| **Parameter** | **Q2D radial-azimuthal** (Xe, Kr, Ar) | **1D radial** (Xe) |
| **Computational parameters** | | |
| Domain's radial length ($L_x$) | 1.28 [cm] | 1.28 [cm] |
| Domain's azimuthal length ($L_z$) | 1.28 [cm] | --- |
| Domain's virtual axial length ($L_y$) | 1 [cm] | 1 [cm] |
| Radial cell size ($\Delta x$) [$\theta$ = -15, -30, 0, +15 to +60] | 50 [$\mu m$] | 50 [$\mu m$] |
| Radial cell size ($\Delta x$) [$\theta$ = -60, -45] | 25 [$\mu m$] | 25 [$\mu m$] |
| Azimuthal cell size ($\Delta z$) [$\theta$ = -15, -30, 0, +15 to +60] | 50 [$\mu m$] | --- |
| Azimuthal cell size ($\Delta z$) [$\theta$ = -60, -45] | 25 [$\mu m$] | --- |
| Time step ($ts$) [$\theta$ = -15, -30, 0, +15 to +60] | $1.5 \times 10^{-11}$ [s] | $1.5 \times 10^{-11}$ [s] |
| Time step ($ts$) [$\theta$ = -60, -45] | $7.5 \times 10^{-12}$ [s] | $7.5 \times 10^{-12}$ [s] |
| Initial number of macroparticles per cell ($N_{ppc}$) | 100 | 100 |
| Total simulated time | 30 [$\mu s$] | 30 [$\mu s$] |
| **Physical parameters** | | |
| Initial plasma number density ($n_{i,0}$) | $1.5 \times 10^{16}$ [m$^{-3}$] | $1.5 \times 10^{16}$ [m$^{-3}$] |
| Initial electron temperature ($T_{e,0}$) | 10 [$eV$] | 10 [$eV$] |
| Initial ion temperature ($T_{i,0}$) | 0.5 [$eV$] | 0.5 [$eV$] |
| Axial electric field ($E_y$) | 10,000 [Vm$^{-1}$] | 10,000 [Vm$^{-1}$] |
| Radial magnetic field intensity at the mid radial plane ($B_x$) | 0.02 [T] | 0.02 [T] |
| Electric potential at the walls ($\phi_w$) | 0 [V] | 0 [V] |

Table 1: Summary of the computational and physical parameters for various Q2D radial-azimuthal and 1D radial simulations

For the negative-curvature simulations with $\theta$ = -45 and -60 degrees, since the steady-state plasma density that establishes in the presence of these large negative curvatures was seen to increase to relatively high values, the cell size and the time step of the corresponding simulations were adapted accordingly to remain below the Debye length and the electron plasma frequency limits, which are associated with the stability of the explicit PIC simulations [32][33]. Therefore, in the simulations for negative curvatures of -45 and -60 degrees, the cell size (cell number) and the time step are, respectively, 25 $\mu m$ (512 nodes) and $7.5 \times 10^{-12}$ s. The total simulated time for all simulations is 30 $\mu s$. Table 1 summarizes the computational and physical parameters used for various simulations in this work.

Initialization is performed by loading uniformly distributed ion-electron pairs at a density of $1.5 \times 10^{16}$ $m^{-3}$. The initial number of macroparticle per cell is 100. The electrons and ions are sampled from Maxwellian distribution functions at the temperatures of 10 eV and 0.5 eV, respectively.

The simulations are collisionless, and the establishment of a steady-state density in the system is ensured through imposing an ionization source, as per the approach of Ref. [15]. The electron-ion pairs are injected at each time step according to this ionization source. The source has an azimuthally uniform distribution and a cosine-shaped profile along the radial direction that spans over $x$ = 0.09 cm to $x$ = 1.19 cm and features a peak value of $8.9 \times 10^{22}$ $m^{-3}s^{-1}$ [15].

A periodic and a zero-volt Dirichlet boundary conditions are applied on the plasma potential along the azimuthal and radial directions, respectively. Secondary electron emission is not accounted for and all particles reaching the walls are removed. Particles crossing the boundaries of domain along the azimuthal direction re-enter the domain from the other end while maintaining their velocities and radial position. We have assumed a finite extent of $L_y$ = 1 cm for the domain along the axial direction in order to prevent the particles' energy to increase indefinitely [34]. The electrons and ions that traverse the artificial axial boundaries of the domain are reintroduced onto the simulation plane, retaining their radial and azimuthal positions and with new velocities sampled from their respective initial Maxwellian distributions.

**Section 3: Results and discussion**

The results of the analyses regarding the impact of the magnetic field curvature ($\theta$) on different aspects of the dynamics of the plasma in the radial-azimuthal configuration will be presented throughout this section. We start by assessing the variation in the radial and 2D distributions of the plasma properties. We then discuss how different degrees of the field's curvature change the characteristics of the developed instabilities, followed by presenting the variations in the electrons' axial mobility. The contribution of different mechanisms to the cross-field electrons' transport will be compared across the simulated range of $B$-field curvatures. Finally, we assess the influence that the field's curvature and the consequent variation in the instability spectra have on the radial and azimuthal divergence of the ion beam. Wherever applicable, we compare the outcomes of the Q2D simulations with the corresponding results obtained from the 1D simulations. Throughout the discussions, we highlight the notable differences in the plasma behavior observed among Xe, Kr, and Ar propellants.

**3.1. Distribution of time-averaged plasma properties for different $B$-field's curvature**

The radial profiles of the ion number density ($n_i$), averaged over the time interval of 20-30 $\mu s$, as well as the variation with $\theta$ of the peak value of these profiles are presented in Figure 3. The number density increases notably as $\theta$ approaches large negative values. However, it remains nearly identical for cases with $\theta$ = -15, 0, -30 degrees. In contrast, $n_i$ exhibits a more moderate variation across the positive values of $\theta$. The density increases from $\theta$ = 0 to $\theta$ = 15 degrees after which it drops with further increase in $\theta$. The general trend of $n_i$ vs $\theta$ is consistent across all propellants, with Xe exhibiting the highest density, Ar displaying the lowest, and Kr falling in between. However, there is an exception in the scenario with $\theta$ = -60 degrees, where Kr demonstrates the highest density among the three propellants.

The normalized ion density profiles ($n_i/n_{i,mean}$) from the Q2D simulations are compared against the corresponding profiles from the 1D simulations in Figure 4. In addition, in Figure 5, the variation vs $\theta$ of the peak of these profiles at the domain's centerline from the 1D and the Q2D simulations is provided.

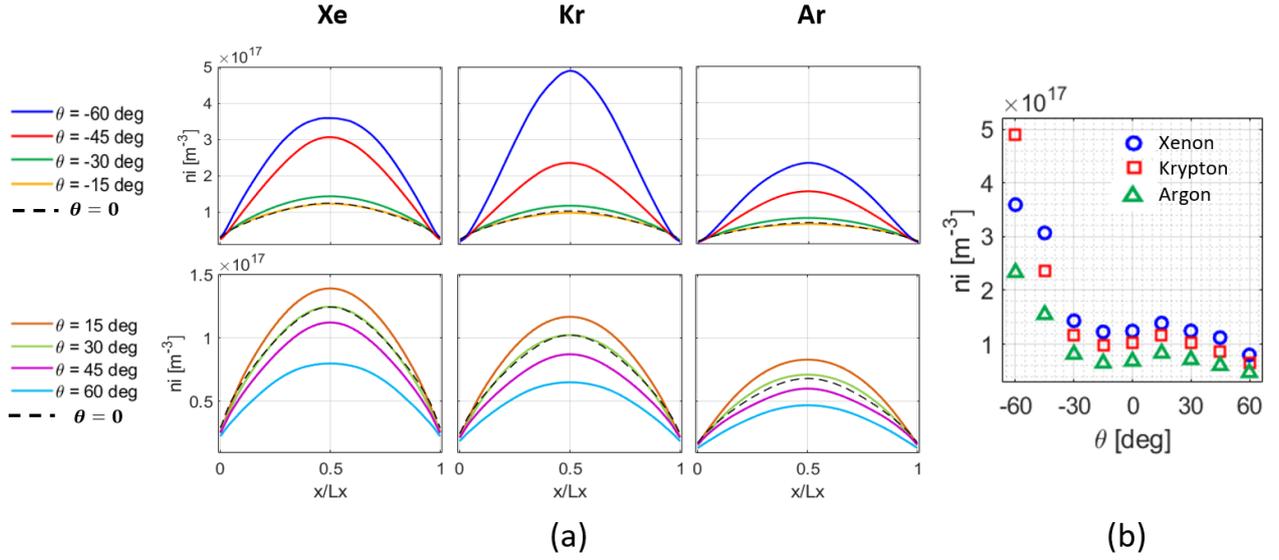

Figure 3: (a) Radial profiles of the ion number density ($n_i$) averaged over 20-30 $\mu s$ from the simulations with various magnetic field curvatures and propellants; (top row) negative curvatures, (bottom row) positive curvatures. (b) Variation vs curvature degree of the ion number density's peak value along the radial direction for the three studied propellants.

It is observed in these plots that the impact of a negative curvature of the $B$-field is more pronounced on the density profile in 1D simulations compared to the Q2D simulations. In the 1D simulations, the density becomes progressively more confined toward the center as the angle $\theta$ changes from 0 to -60 degrees. However, the concentration effect of the negative curvature of the $B$-field is mitigated in the Q2D simulations due to presence of the azimuthal instabilities and their consequences on the plasma properties which will be discussed later.

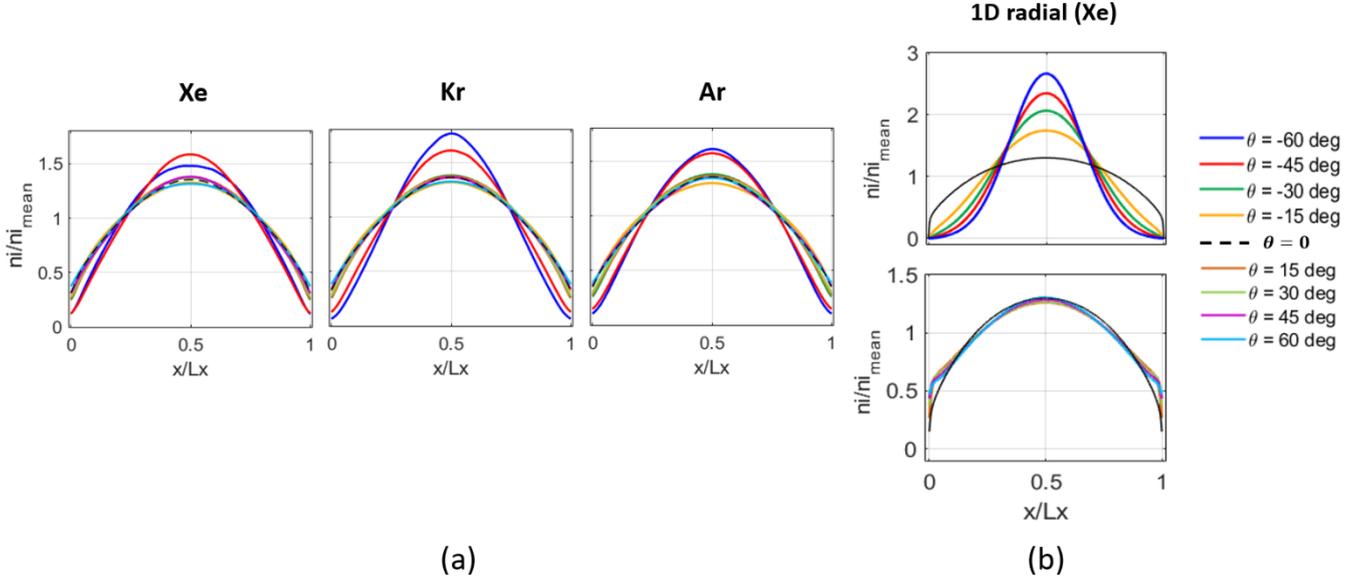

Figure 4: (a) Time-averaged radial profiles of the ion number density normalized with the radially averaged value of the ion density ($n_{i,mean}$) from the radial-azimuthal simulations with various magnetic field curvatures and propellants. (b) Normalized ion number density radial profiles from the reference 1D radial simulations with negative (top plot) and positive (bottom plot) curvatures and with xenon propellant.

In the Q2D simulations, the concentration of plasma toward the center is most notable for the cases with $\theta = -60$ and -45 degrees. Moreover, for $\theta = -60$ degrees, the extent of the confinement is different among the three propellants. Apart from these observations, the $n_i/n_{i,mean}$ profiles are rather similar across the rest of the curvatures ($\theta = -30$ to 60 degrees) and propellants. Furthermore, unlike the negative curvatures, the positive curvatures of the $B$-field are noticed to have a minor influence on the normalized density profiles from the 1D simulations, hence, leading to a better agreement with the $n_i/n_{i,mean}$ profiles from the Q2D simulations.

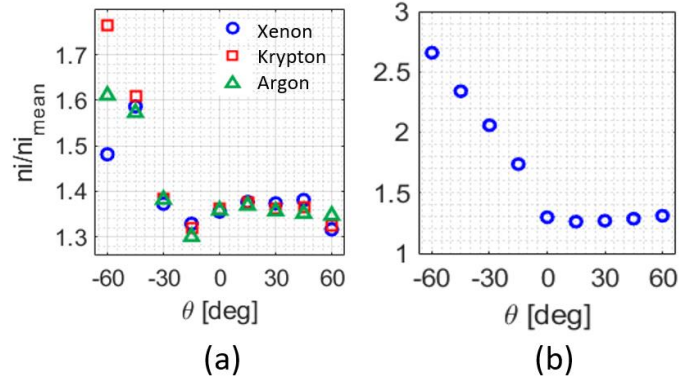

Figure 5: (a) Variation vs $\theta$ of the normalized ion number density's peak value along the radial direction for the three studied propellants from the quasi-2D simulations. (b) Variation vs $\theta$ of the normalized ion number density's peak value from the reference 1D radial simulation for the xenon propellant.

We now look at the time-averaged profiles of the radial electron temperature ($T_{ex}$) in Figure along with the variation vs $\theta$ of the mean of the $T_{ex}$ profiles in Figure 6.

It is noticed that the $T_{ex}$ is much higher in the Q2D simulations compared to the 1D simulations due to the electrons' heating effect of the instabilities captured in the Q2D simulations. Moreover, in the presence of negative curvatures, in the 1D simulations, $T_{ex}$ monotonically increases as the $\theta$ varies from zero toward more negative values. Contrarily, in the Q2D setting, larger degrees of curvature ($\theta = -60$ and $-45$ degrees) lead to lower $T_{ex}$ values compared to the baseline case ($\theta = 0$), whereas moderate levels of curvature ($\theta = -30$ and $-15$ degrees) cause $T_{ex}$ to be higher than that in the baseline case. The disparity in the trend observed in the density profiles for Kr compared to the other propellants at $\theta = -60$ degrees is evident in the behavior of $T_{ex}$ as well.

The positive curvatures of the $B$-field flatten the distribution of the radial electron temperature which leads to more uniform profiles across the radial direction both in the 1D and the Q2D simulations. The dependency of $T_{ex}$ on the field's curvature is also minor across positive values of $\theta$.

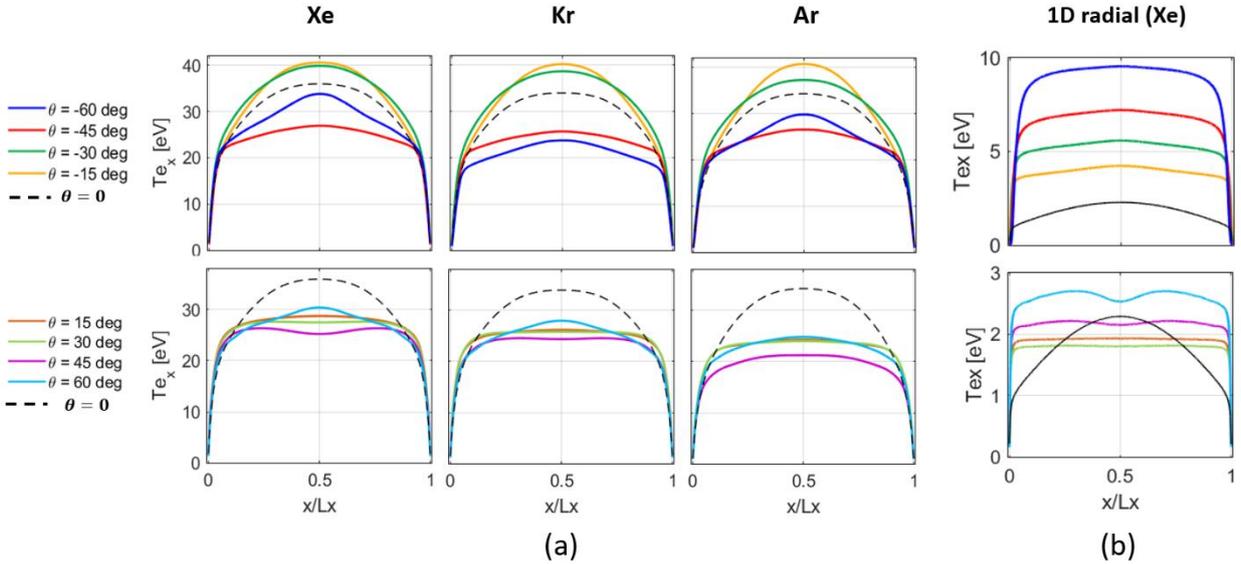

Figure: (a) Profiles of the radial electron temperature ($T_{ex}$) averaged over 20-30 $\mu s$ from the quasi-2D simulations with various magnetic field curvatures and propellants; (top row) negative curvatures, (bottom row) positive curvatures. (b) Time-averaged radial profiles of $T_{ex}$ from the reference 1D radial simulations for xenon propellant and negative (top plot) and positive (bottom plot) $B$-field curvatures.

To summarize the overall trend of the $T_{ex}$ from the Q2D simulations, it is seen in Figure 6 that the radial electron temperature monotonically decreases with $\theta$ from $-30$ to $45$ degrees and it increases slightly at $\theta = 60$ degrees. Across this range of B-field curvatures, Ar consistently demonstrates the lowest $T_{ex}$, whereas Xe exhibits the highest values.

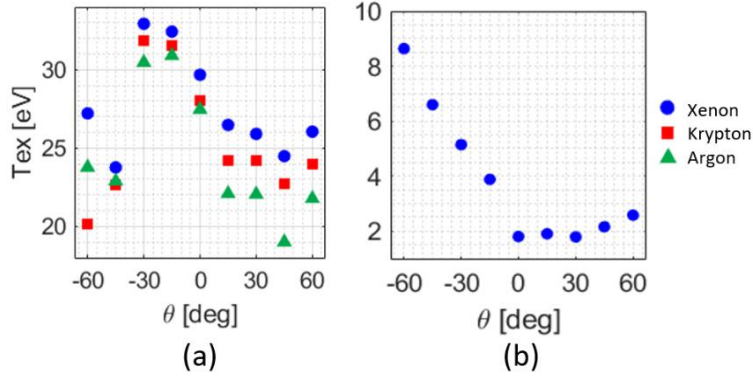

Figure 6: (a) Variation vs $\theta$ of the mean radial electron temperature value along the radial direction for the three studied propellants from the quasi-2D simulations. (b) Variation vs $\theta$ of the mean $T_{ex}$ value from the reference 1D radial simulation for xenon propellant.

To assess potential anisotropy in the electron temperature and its variations with the curvature degree, we have plotted in Figure 7(a), the average (over the entire domain and over the time interval of $20-30\ \mu s$) radial-to-azimuthal electron temperature ratio ($T_{ex}/T_{ez}$) against $\theta$ for the Q2D simulations. Based on the results reported in Refs. [35][36] regarding the contributions that the Electron Cyclotron Drift instability (ECDI) [37][38] and the Modified Two Stream Instability (MTSI) [16][39] have to the electrons' heating along the azimuthal and the radial directions, respectively, the $T_{ex}/T_{ez}$ ratio has been showed to be an indicative measure of the relative dominance of these two instability modes. Accordingly, two regimes can be defined: 1) $\frac{T_{ex}}{T_{ez}} > 1$, which signifies the prevalence of the MTSI, and 2) $\frac{T_{ex}}{T_{ez}} < 1$, which indicates that the ECDI is the dominant instability.

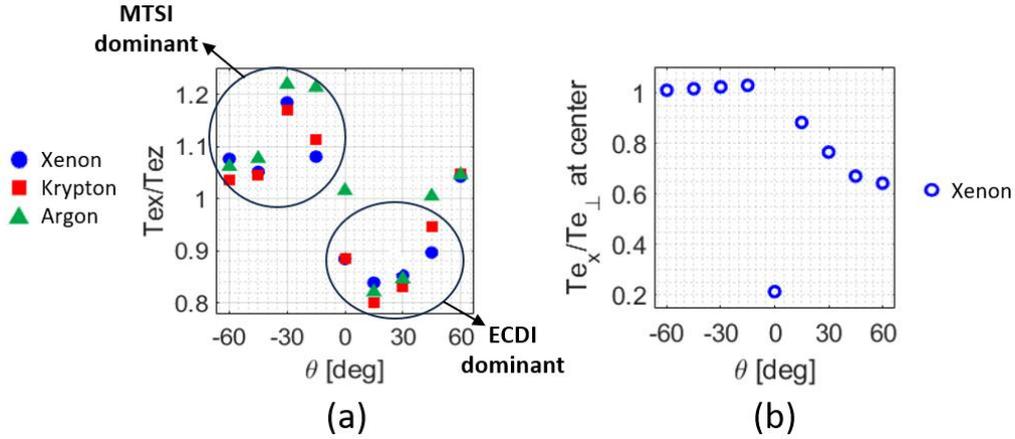

Figure 7: (a) Variation vs $\theta$ of the radially averaged values of $T_{ex}/T_{ez}$ from the quasi-2D simulations with various magnetic field curvatures and different propellants. (b) Variation vs $\theta$ of the temperature ratio $T_{ex}/T_{e,\perp}$ at the mid-radial point for the reference 1D radial simulation with xenon propellant.

Referring to Figure 7(a), we can distinctly notice the two regimes mentioned above. Indeed, the presence of negative B-field curvatures results in $\frac{T_{ex}}{T_{ez}} > 1$ and, therefore, the dominance of the MTSI. Conversely, the positive curvatures lead, in general, to $\frac{T_{ex}}{T_{ez}} < 1$, which is attributed to the ECDI modes of relatively high strength. A particular exception to these general observations are noticed for the case of $\theta = 60$ degrees where the $\frac{T_{ex}}{T_{ez}}$ ratio is slightly higher than 1 for all propellants.

There are also cases where different propellants exhibit significantly disparate anisotropy, namely, the cases with $\theta$ = -15, 0, and 45 degrees. In each of these scenarios, the $\frac{T_{ex}}{T_{ez}}$ ratio is the highest for Ar which indicates the presence of noticeably stronger MTSI modes compared to the simulations with the other two propellants. Furthermore, the cases where $\frac{T_{ex}}{T_{ez}} \sim 1$, i.e., $\theta = 0$ and 45 with Ar, suggest the coexistence of the ECDI and the MTSI modes with comparable intensities.

To verify the above arguments, we have conducted a detailed assessment of the instabilities' characteristics, which will be presented later in Section 3.2. The findings from the instability analyses have been in general consistent with what we observed here, attesting the applicability of the $\frac{T_{ex}}{T_{ez}}$ ratio as a representative criterion to assess the comparative significance of the ECDI and the MTSI.

In the 1D simulations, we defined the temperature anisotropy as the ratio of the parallel-to-perpendicular electron temperature ($Te_\parallel/Te_\perp$), where the parallel and perpendicular directions are with respect to the magnetic field. The variation with $\theta$ of $Te_\parallel/Te_\perp$ at the centerline from the 1D simulations is provided in Figure 7(b). It is noted that, at the centerline, the magnetic field aligns with the radial direction, resulting in $Te_\parallel$ to be equivalent to $T_{ex}$. We notice from Figure 7(b) that introducing any level of B-field curvature significantly enhances the temperature isotropy in the system, which is in line with the observed behavior in Ref. [24]. In fact, for a purely radial $B$-field ($\theta = 0$), the $T_{ex}/Te_\perp \approx 0.2$. However, this ratio shifts to around 1 in the presence of the negative curvature in the magnetic field and stays almost constant across $\theta = $ -60 to -15 degrees. For the positive curvatures, the $T_{ex}/Te_\perp$ ratio decreases as $\theta$ increases, even though it still maintains significantly higher values when compared to the scenario of no curvature.

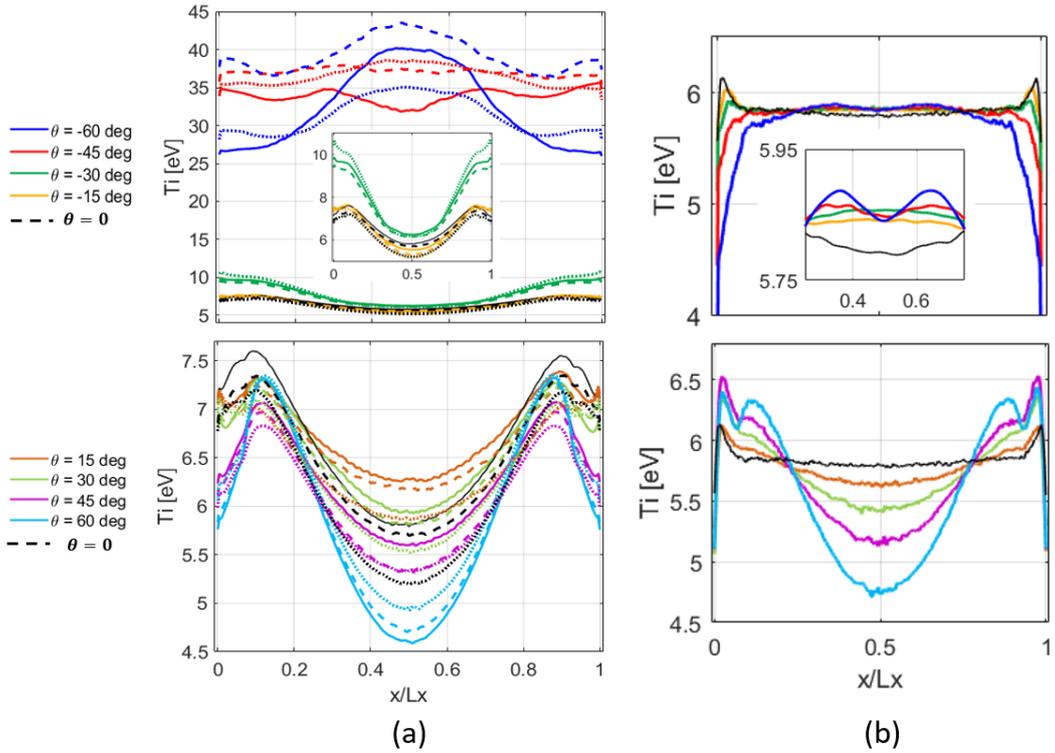

Figure 8: (a) Radial profiles of the ion temperature ($T_i$) averaged over 20-30 $\mu s$ from the quasi-2D simulations with various magnetic field curvatures and propellants: solid lines correspond to xenon, dashed lines to krypton, and dotted lines to argon; (top row) negative curvatures, (bottom row) positive curvatures. (b) Time-averaged radial profiles of $T_i$ from the 1D radial simulations for xenon propellant and negative (top plot) and positive (bottom plot) $B$-field curvatures.

In Figure 8, we compare the time-averaged profiles of the ion temperature ($T_i$) for various $\theta$ from the Q2D and the 1D simulations. From the plots in Figure 8 (a), in the Q2D simulations with large negative $B$-field curvatures ($\theta = $ -45 and -60 degrees), the ion temperature reaches up to 35-45 eV. As reported in Refs. [26][29], such high $T_i$ values occur as a result of strong interactions between the ion particles and an instability mode characterized by a high frequency and a relatively long wavelength, which we had called the "long-$\lambda$" mode in Ref. [26]. We will demonstrate later from the instability analyses' results (Section 3.2) that this mode indeed appears under the large negative curvature conditions.

We also observe from the Q2D simulations' results in Figure 8 that, at $\theta = $ -60 degrees, $T_i$ is quite higher for Kr propellant than for Xe and Ar. For $\theta$ values of -30 to 0 degrees and for all positive curvatures, the ion temperature is lower than the values observed for the $\theta = $ -60 and -45 degrees. The radial distribution of the $T_i$ is noticed to be varying across the studied $\theta$ range. In particular, at $\theta = $ -60 degrees, $T_i$ is higher at the centerline. For $\theta = $ -45 degree, the $T_i$ profile is relatively uniform over the domain, and for $\theta = $ -30 to 0 degrees, the $T_i$ peaks near the walls. On the contrary, for all positive curvatures, the shape of the profiles is seen to be similar, with the ion temperature peaking near the wall and dropping toward the center of the domain.

In contrast to the Q2D simulations, in the 1D setting, under negative *B*-field curvature conditions, the $T_i$ does not substantially vary with $\theta$ in the bulk plasma. However, a minor reduction in $T_i$ values is observed adjacent to the walls as $\theta$ changes from 0 to -60 degrees. In the cases with positive curvatures, the $T_i$ profiles obtained from the 1D and the Q2D simulations are more consistent. In these cases, as $\theta$ increases from 0 to 60 degrees, the $T_i$ profile becomes progressively less uniform with a consistent reduction in values at the centerline.

Figure 9 shows the time-averaged radial distribution of the ion Mach number from the Q2D and the 1D simulations for various *B*-field curvatures. Interestingly, in the Q2D simulations of all propellants, as $\theta$ increases from -60 to 60 degrees, the ion sonic point continuously progresses toward the bulk plasma away from the walls. Contrary to the Q2D results, in the 1D simulations of the xenon propellant, the location of the sonic point slowly gets closer to the walls when $\theta$ varies from -60 degrees to 0, whereas it moves rapidly away from the walls as $\theta$ increases for the positive curvatures.

Our observation from the 1D simulations has also been reported in Ref. [24], where the authors noticed a similar trend with the field's curvature. The general behavior of the sonic point from Ref. [24] is, however, only in line with the Q2D simulations' results for the positive *B*-field curvatures. Even in this case, the sensitivity of the position of the ion sonic point to the curvature of the magnetic field is still more significant in a 1D setting.

The occurrence of the sonic condition for ions within the plasma bulk, where the quasi-neutrality holds, renders the Bohm's criterion inadequate in establishing the boundary (edge) of the sheath. Thus, as also suggested in Ref. [24], in these situations, the non-quasineutrality serves instead as the appropriate criterion to determine the sheath's edge. According to this criterion [24], the sheath's boundary is defined to coincide with the location where the non-quasineutrality exceeds a certain threshold, typically beyond 5% of the bulk ion density, i.e.,

$$\frac{|n_i - n_e|}{n_i} \approx 0.05 . \tag{Eq. 2}$$

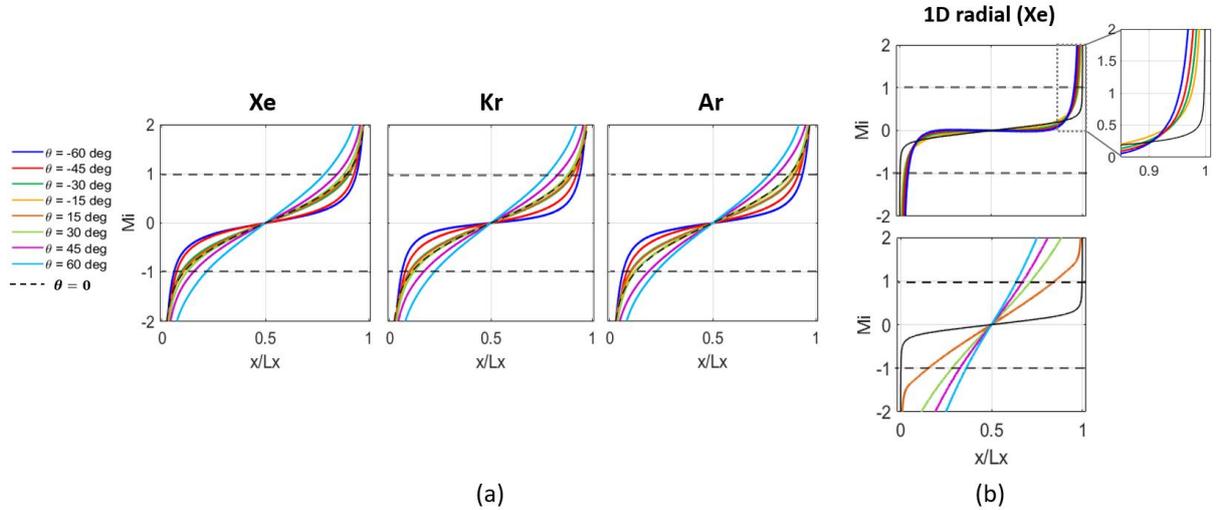

Figure 9: (a) Time-averaged radial profiles of the ion Mach number over the time interval of 20 – 30 $\mu s$ from the quasi-2D simulations with various magnetic field curvatures and propellants. (b) Ion Mach number's radial profiles from the 1D radial simulations with negative (top plot) and positive (bottom plot) curvatures and with xenon propellant.

Figure 10 compares, for the Q2D and 1D simulations and for different *B*-field curvatures, the locations of the ion sonic point and the sheath' edge as described by the non-quasineutrality condition (Eq. 2) in the vicinity of the right-hand-side wall of the domain. According to Figure 10(b), in the 1D simulations and across the negative curvatures, the Bohm's criterion agrees with the non-quasineutrality condition in determining the location of the sheath's edge. However, in the cases of positive curvature, there is a notable disparity between the two criteria. In the Q2D setting, we see from Figure 10(a) that, apart from the positive-curvature cases, the disagreement of the criteria extends to the negative curvatures as well including the cases with $\theta$ = -15 and -30 degrees.

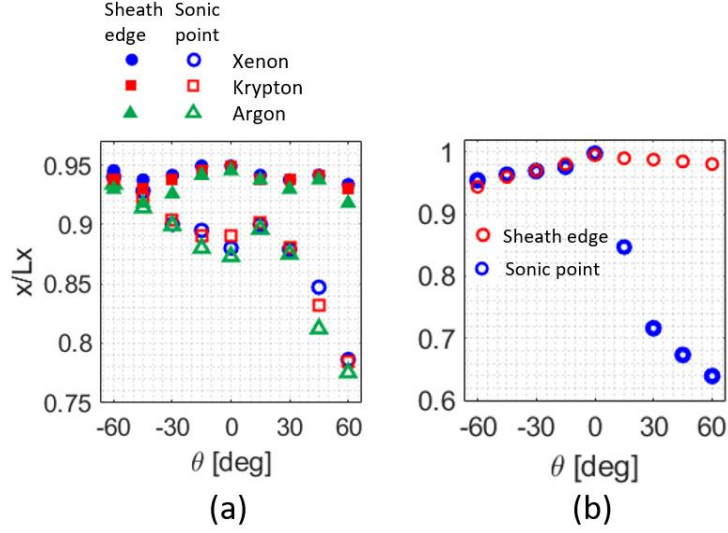

Figure 10: Variation vs $\theta$ of the normalized radial location of the ion sonic point and the sheath edge (from Eq. 2) for the quasi-2D simulation with various propellants. (b) Similar analysis for the 1D radial simulation with xenon propellant.

Finally, before presenting the results of the spectral analyses of the instabilities, looking at the 2D radial-azimuthal snapshots of the plasma properties in Figure 11 and Figure 12 allows us to gain a comparative perspective on the dominant instabilities.

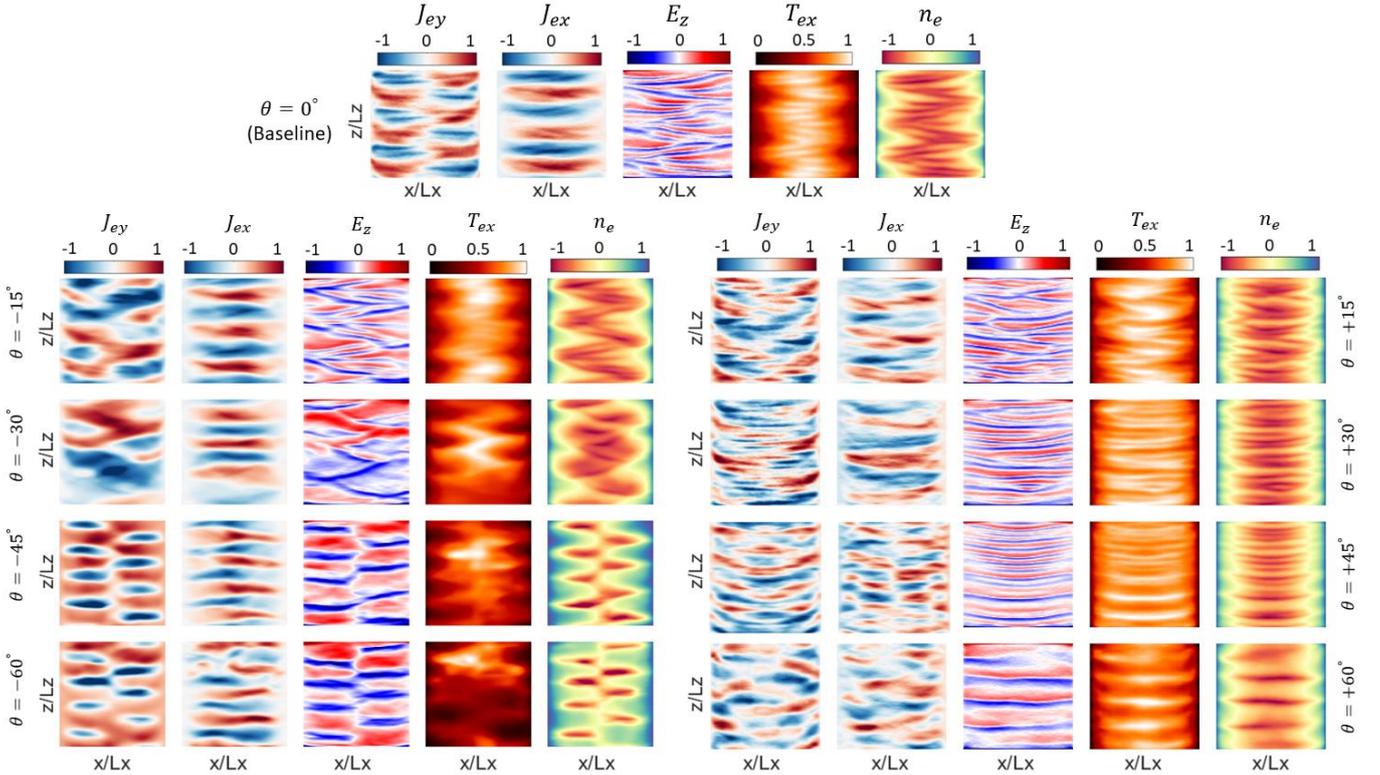

Figure 11: Comparison of the 2D snapshots of the normalized plasma properties at a time of local maximum of the radial electron temperature for various $B$-field curvatures and the xenon propellant. The columns, from left to right, represent the axial and radial electron current densities ($J_{ey}$ and $J_{ex}$), the azimuthal electric field ($E_z$), the radial electron temperature ($T_{ex}$), and electron number density ($n_e$).

Figure 11 illustrates the snapshots of various plasma properties from the Q2D simulations with varying $\theta$ values and for the Xe propellant. The snapshots are captured at a time instant corresponding to a local peak of the radial electron temperature ($T_{ex}$), during which the MTSI, if present, is expected to be strong [15][28]. It is evident from this figure that the negative- and the positive-curvature cases display distinctively different behaviors in terms of the spatial structures of the involved fluctuations. In general, the negative curvature of the B-field tends to reinforce the development of the MTSI-representative patterns across the plasma properties, whereas the positive curvatures seem to suppress the MTSI, leaving the azimuthal waves associated with the ECDI as the primary instability modes in the system. The absence of the B-field curvature ($\theta = 0$) represents an intermediate scenario, where the patterns resembling both the ECDI and the MTSI are quite visible in the snapshots. Moreover, the appearance of relatively long-wavelength structures can be noticed in the cases with $\theta$ = -60 and -45 degrees.

Figure 12 depicts a comparison of the snapshots of the axial electron current density ($J_{ey}$) and the azimuthal electric field ($E_z$) between Xe, Kr, and Ar at their respective instances of the local maximum of $T_{ex}$ and for different magnetic field curvatures. These sample snapshots show an overall similarity of the distributions for the three propellants. Nevertheless, a rigorous instability characterization enables distinguishing the detailed differences that exist among various propellants. This will be presented in the next section.

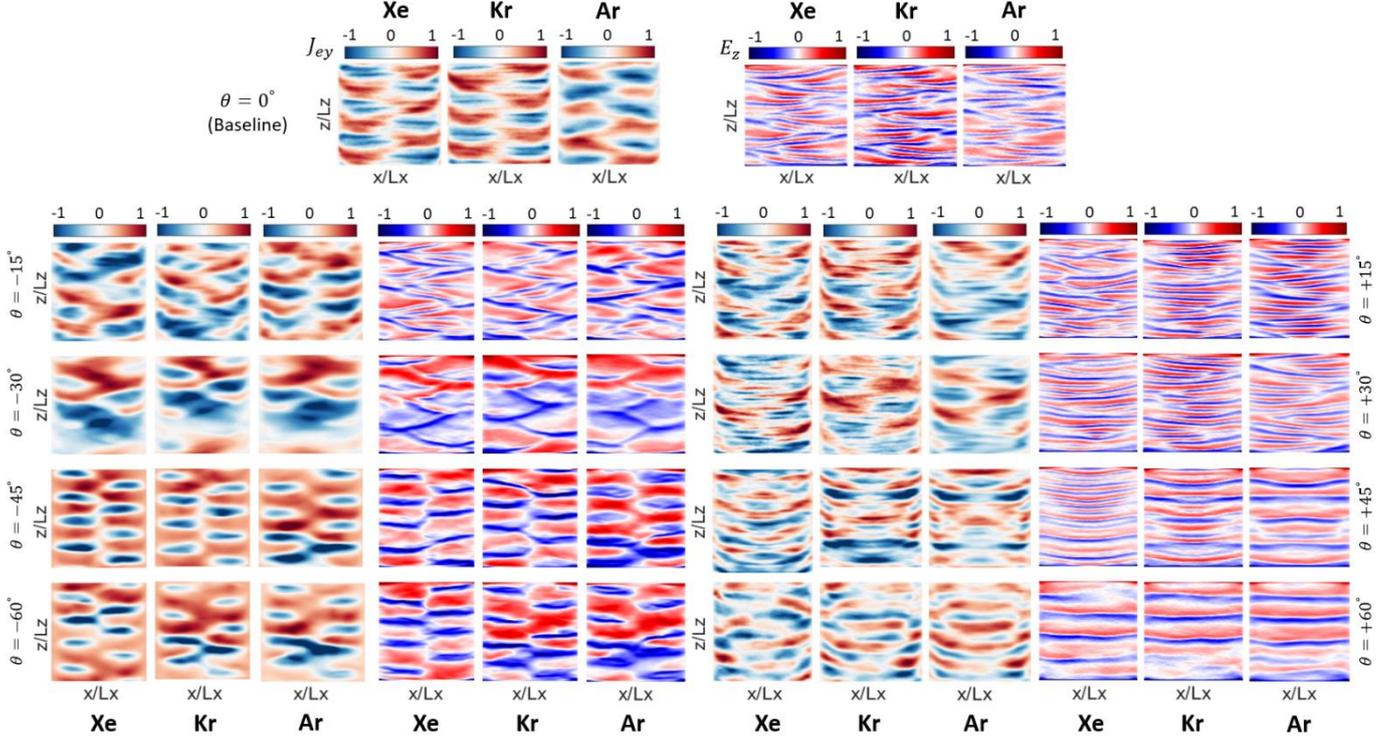

Figure 12: Comparison of the 2D snapshots of the normalized radial electron current density ($J_{ey}$) and the normalized azimuthal electric field ($E_z$) at a time of local maximum of the radial electron temperature for various B-field curvatures and different propellants.

### 3.2. Variation in the spectra and structure of the azimuthal instabilities

We carried out spectral analyses of the azimuthal fluctuations from the Q2D simulations using the spatial and temporal Fast Fourier Transform (FFT) as well as the Dynamic Mode Decomposition (DMD) [40] to study the instabilities' characteristics, including the wavenumber, frequency, and spatial structure, in addition to the variations of these across different $\theta$ values and with various propellants. The analyses are performed on the spatiotemporal signal of the azimuthal electric field ($E_z$). We report the results of the spectral analyses separately for the negative- and positive-curvature simulations.

Starting with the cases with negative B-field curvature, Figure 13 shows the spatial FFT of the $E_z$ averaged across all radial positions and over the time span of 20-30 $\mu s$. In these plots, the horizontal axis represents the wavenumber which is normalized with respect to the fundamental resonance wavenumber of the ECDI, $k_0 = \frac{\Omega_{ce}}{V_{d_e}}$, where $\Omega_{ce}$ and $V_{d_e}$ are the electron cyclotron frequency and the electron azimuthal drift velocity, respectively. We notice from this figure that the ECDI is developed only in cases where $\theta$ = 0 or -15 degrees. Interestingly, in these cases, the intensity of the ECDI is lower for Ar compared to Xe and Kr. The higher curvatures of the magnetic field suppress the ECDI. Indeed, in the cases with $\theta$ = -60, -45 and -30 degrees, the ECDI is absent. Moreover, the second harmonics of the ECDI are not observed in any of the cases except for the baseline ($\theta = 0$).

The MTSI and the long-$\lambda$ instability [26] are observed in all negative $\theta$ cases, with the amplitudes that are nearly the same across all propellants. The intensity of the long-$\lambda$ instability at $\theta$ = -15 degrees is lower than the other cases. This instability is significantly damped at $\theta = 0$ for Xe and Kr, whereas in the case of Ar, its intensity is notably higher. As pointed out in Section 1, the long-$\lambda$ instability has been observed in our previous work [26] in the presence of strong magnetic field gradients where it was found to have remarkable impacts on the plasma processes. These included significant electron cross-field transport and strong interaction with the ions leading to large ions' azimuthal drift and heating [26]. We will show later that these effects are observed in the present simulations as well.

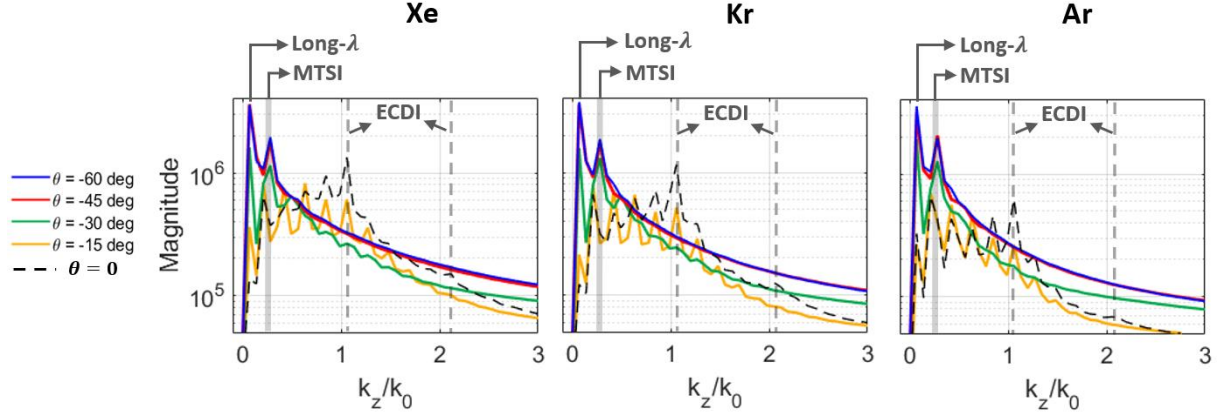

Figure 13: 1D spatial FFT plots of the azimuthal electric field signal from simulations with various negative magnetic field curvatures and different propellants. The FFTs are averaged over all radial positions and over the time interval of 20-30 $\mu s$.

To assess the frequency spectra of the instabilities, the averaged temporal FFTs of the $E_z$ for negative-curvature simulations of Xe, Kr, and Ar are provided in Figure 14. Looking at these plots, the peaks below 10 MHz in the frequency spectrum consistently shift toward higher frequencies as the propellant changes from Xe to Kr to Ar. Also, the frequency spectrum of Ar in cases of $\theta = 0$, -15 and -30 degrees exhibits more distinct peaks below 10 MHz compared to the Kr and Xe simulations. Another noteworthy observation is that, at $\theta$ = -30 degrees, the FFT plots reveal the presence of instability modes with frequencies in the range of 20-25 MHz. The amplitude of these modes decreases from Xe to Kr, and they are nearly suppressed in the case of Ar.

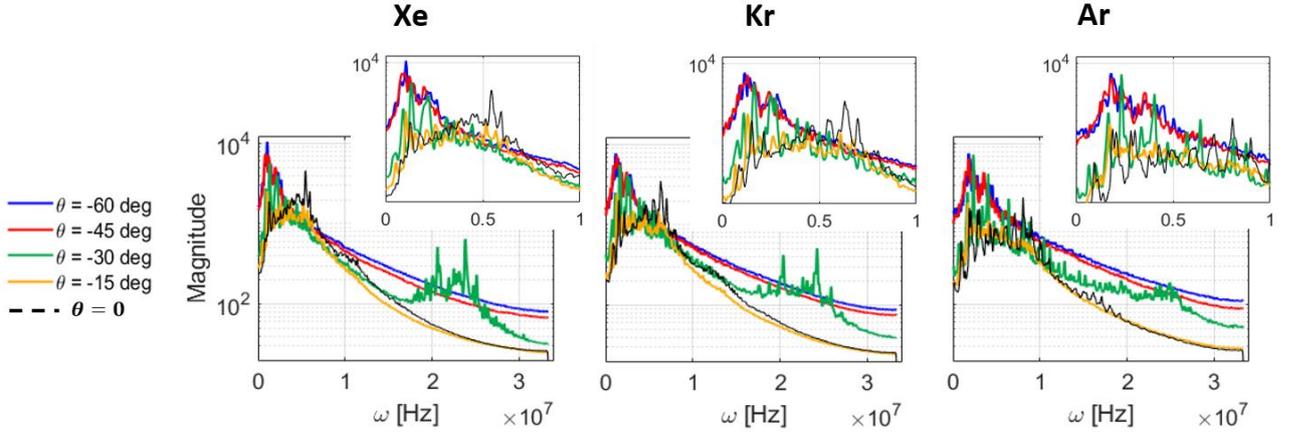

Figure 14: 1D temporal FFT plots of the azimuthal electric field signal from simulations with various negative magnetic field curvatures and different propellants. The FFTs are averaged over all radial positions. The zoomed-in view on each FFT plot in the frequency range of 0 – 10 MHz is also shown as an embedded subplot.

To understand the spatial structure of the instability modes corresponding to the peaks in the frequency spectra, we applied the DMD to the time series 2D snapshots of the $E_z$. Unlike the spatial FFT, which operates under the assumption of a strictly harmonic foundation in spatial representation, DMD takes a more adaptive approach by utilizing generic spatial bases that are uniquely tailored to the dataset under study. As a result, DMD enables extracting the coherent spatiotemporal modes from the time-series data. These modes are essentially the distinct spatial patterns ($\phi_i$) that capture the dominant structures and behaviors within the data. Additionally, these modes are coupled with their respective linear dynamics over time, which are characterized by complex frequencies ($\omega_i$). The $E_z$ time-series snapshots can be reconstructed following the relation in Eq. 3.

$$E_z(x,z,t) \approx \sum_{i=1}^{r} b_i \phi_i(x,z) \exp(\omega_i t) \,. \tag{Eq. 3}$$

We used Optimized DMD (OPT-DMD), which is a significantly improved variant of the DMD method. In Ref. [41], we demonstrated across several test cases how the robustness of the OPT-DMD enables reliable identification and isolation of stable spatiotemporal modes within the plasma system.

The spatial structure and the frequency of several dominant DMD modes for the Q2D Xe simulations with the negative $B$-field curvatures are presented in Figure 15. The corresponding DMD modes for the baseline case ($\theta = 0$) can be found in Ref. [41].

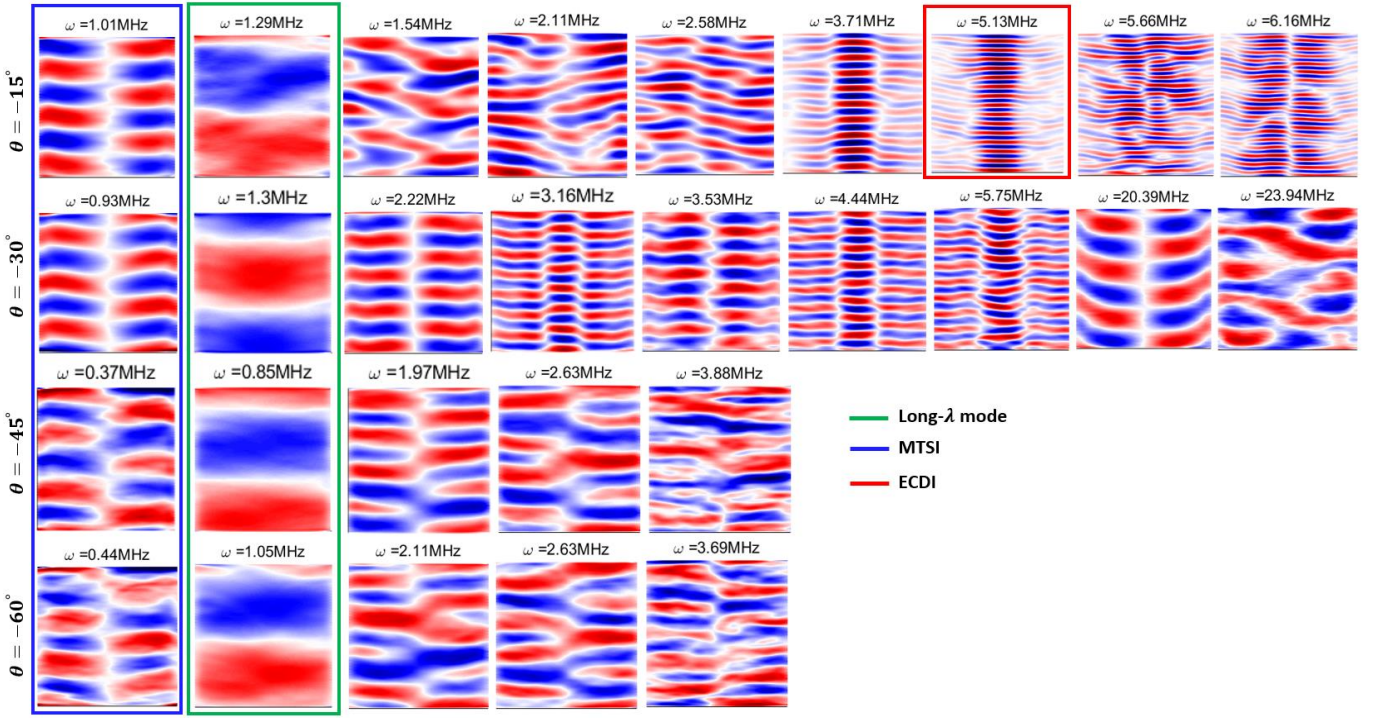

Figure 15: Visualization of the several dominant DMD modes of the azimuthal electric field data from the radial-azimuthal simulations with different negative $\theta$ values and the xenon propellant. The approach pursued to derive these modes are explained in detail in Ref. [41]. The modes identified with a box correspond to those with wavelengths equal or close to the wavelengths of the dominant modes visible in the spatial FFT plot in Figure 13.

The modes enclosed within boxes are those whose wavelength matches the instabilities identified from the spatial FFT plots. Consistent with what we observed from the spatial FFT spectra, the DMD discovers azimuthal waves corresponding to the ECDI only in the case of $\theta = -15$ degrees. Whereas, the MTSI as well as the azimuthal long-$\lambda$ mode with a wavelength approximately equal to the domain's length and a frequency of about 1 MHz are identified in all cases. All the remaining modes detected in various cases contain both radial and azimuthal wavenumbers. In addition, the DMD discovered the spatial configuration of the high-frequency modes that have appeared in the frequency spectrum of the case with $\theta = -30$ at about 20-25 MHz. It turns out that the mode with the frequency of $\sim$ 20 MHz has the same wavelength as the MTSI mode with the frequency of about 1 MHz.

We now proceed with presenting the same set of analyses as those presented so far for the cases with positive $B$-field curvatures. Figure 16 shows the azimuthal wavenumber ($k_z$) spectra of the instabilities obtained from the spatiotemporally averaged spatial FFTs of $E_z$. From this figure, it is evident that the positive curvature of the magnetic field contributes to the suppression of the MTSI, especially for Xe and Kr. In particular, the peaks associated with the MTSI in the $k_z$ spectra of Xe and Kr at $\theta = 15$ and 30 degrees exhibit notably reduced amplitudes in comparison to the baseline case ($\theta = 0$). Whereas, the MTSI's intensity for Ar at these curvature values is comparable to the baseline simulation.

Furthermore, the $k_z$ spectrum of the cases with larger positive $B$-field curvatures, contains peaks at $k_z/k_0 \sim 0.42$ and $k_z/k_0 \sim 0.35$ for the cases with $\theta = 45$ and 60 degrees, respectively. However, it is not readily evident if these peaks correspond to the MTSI since they represent larger azimuthal wavenumbers than those expected from the baseline case ($k_z/k_0 \sim 0.2$). As we will show shortly, the DMD analysis provides further insights into the nature of these modes by characterizing their spatial structures. It will indeed clarify that the peak at $k_z/k_0 \sim 0.35$ for Xe is associated with an MTSI wave.

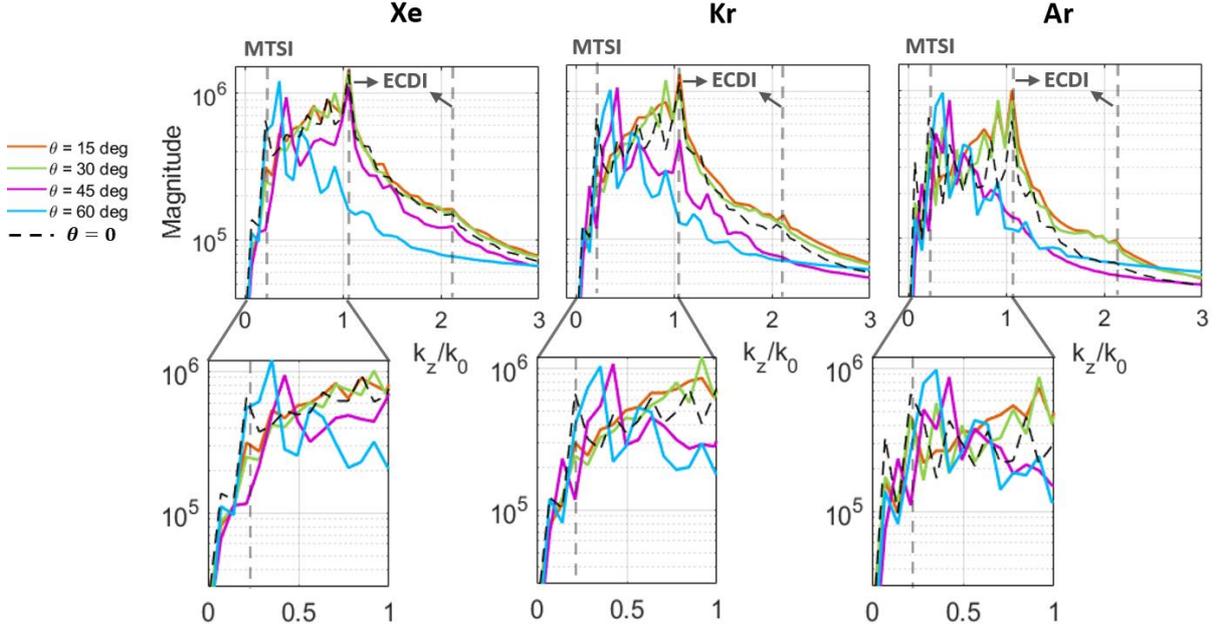

Figure 16: 1D spatial FFT plots of the azimuthal electric field signal from simulations with various positive magnetic field curvatures and different propellants. The FFTs are averaged over all radial positions and over the time interval of 20-30 $\mu s$.

The first harmonic of the ECDI is the dominant mode in the cases of $\theta = 15$ and 30 degrees with the amplitude being slightly lower in the case of Ar than for Xe and Kr. The $\theta = 45$ degrees condition leads to the largest discrepancy among different propellants in terms of the intensity of the ECDI. At this curvature, the ECDI manifests its highest intensity for Xe, while it is mitigated for Kr and is absent in the case of Ar. Notably, in the case of $\theta = 60$ degrees, the nearest peak to the ECDI's first harmonic is shifted toward lower wavenumbers ($k_z/k_0 \sim 0.92$). Additionally, the second harmonic of the ECDI is present in the cases of $\theta = 15$ and 30 degrees for all propellants as well as in the case of $\theta = 45$ degrees for the Xe.

The comparison between the frequency spectra of the $E_z$ from the simulations with the positive $B$-field curvatures is provided in Figure 17. Similar to the trend observed in the negative-curvature cases, the peaks representing the modes below 10 kHz shift toward consistently higher frequencies from Xe to Kr to Ar. It is noteworthy that, in the case of $\theta = 60$ degrees, the FFT spectrum exhibit numerous peaks across a wide range of frequencies.

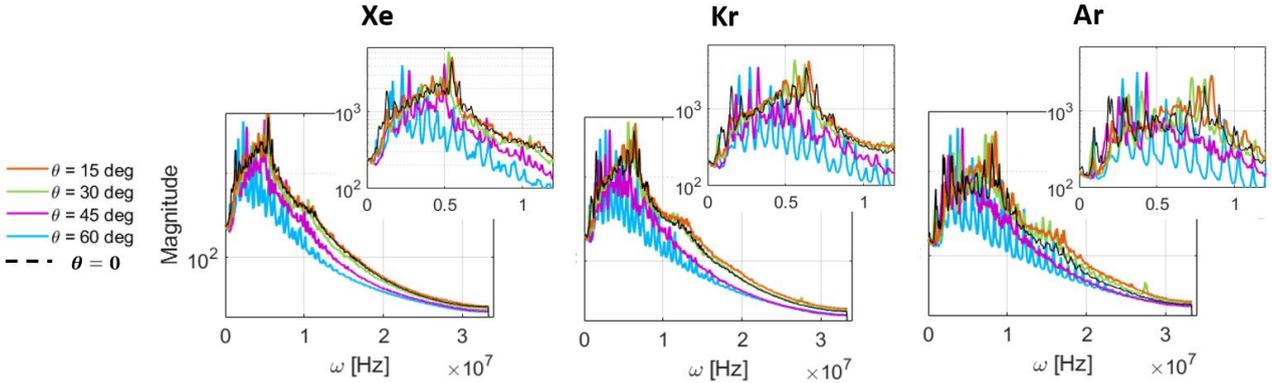

Figure 17: 1D temporal FFT plots of the azimuthal electric field signal from simulations with various positive magnetic field curvatures and different propellants. The FFTs are averaged over all radial positions. The zoomed-in view on each FFT plot in the frequency range of 0 – 10 MHz is also shown as an embedded subplot.

Finally, the dominant spatiotemporal modes obtained from the OPT-DMD analyses of various positive-curvature cases with Xe propellant as an example are illustrated in Figure 18. Most of the presented modes comprise purely azimuthal fluctuation patterns with different wavelengths and frequencies. We notice that, only at $\theta = 60$ degrees, a mode with a radial-azimuthal structure at the frequency of 1.48 MHz is present which could be possibly linked to the MTSI.

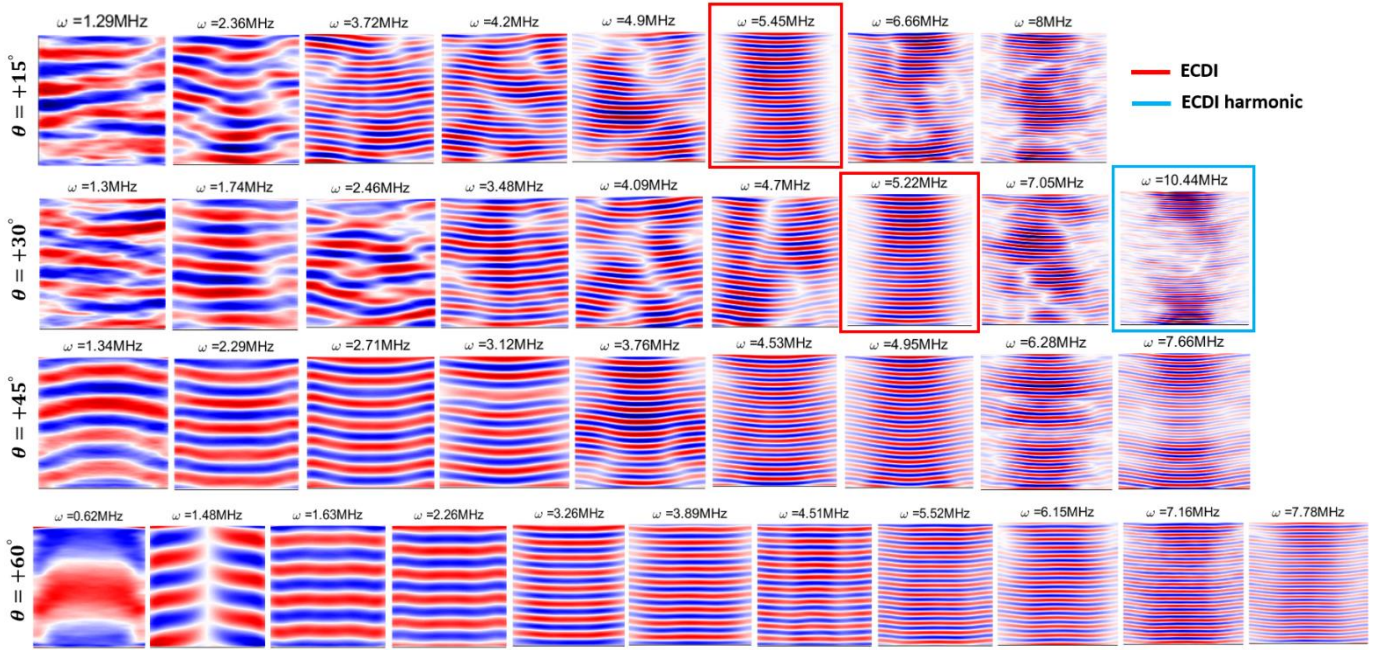

Figure 18: Visualization of the several consecutive dominant DMD modes of the azimuthal electric field data from the radial-azimuthal simulations with different positive $\theta$ values and the xenon propellant. The approach pursued to derive these modes are explained in detail in Ref. [41]. The modes identified with a box correspond to those with wavelengths equal to the wavelengths of the dominant modes visible in the spatial FFT plot in Figure 16.

### 3.3. Influence on the electrons' axial transport

The time averaged (over 20-30 $\mu s$) radial profiles of the electrons' axial mobility ($\mu = \left|\frac{v_{e,y}}{E_y}\right|$) for various $\theta$ values from the Q2D simulations with different propellants and from the 1D simulations with Xe are presented in Figure 19. In the relation for $\mu$, $v_{e,y}$ is the average electrons' axial drift velocity, and $E_y$ is the axial electric field.

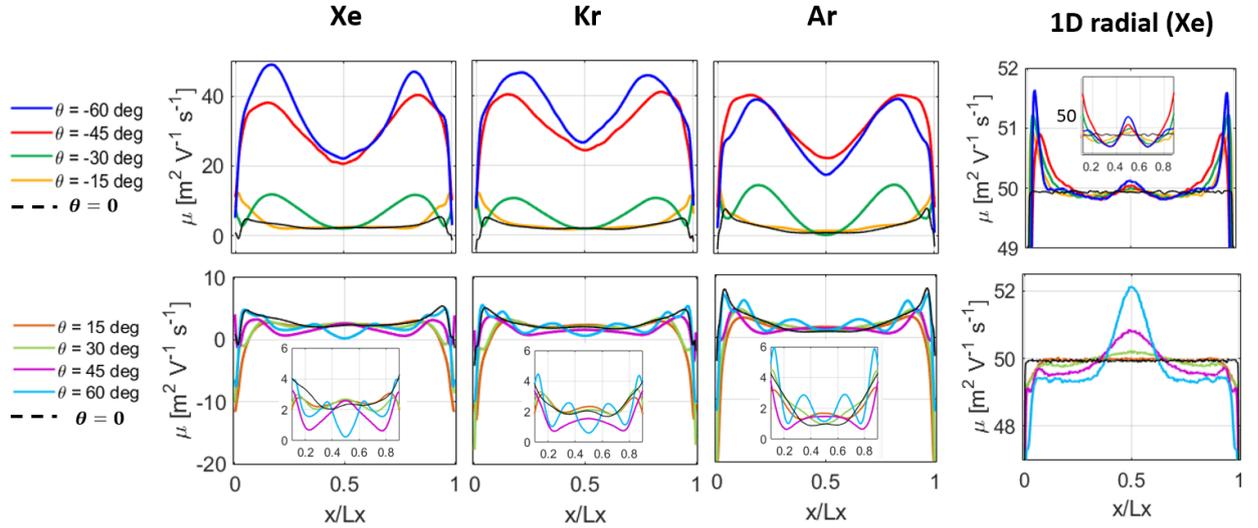

Figure 19: Time-averaged (over 20-30 $\mu s$) radial profiles of electrons' axial mobility ($\mu$) from the quasi-2D simulations with various magnetic field curvatures and propellants. Time-averaged radial profiles of $\mu$ from the 1D radial simulations for xenon propellant and negative (top plot) and positive (bottom plot) $B$-field curvatures are also shown for reference on the right-most column.

Comparing the magnitudes of the electron mobility from the Q2D and the 1D simulations, it is evident that, across all curvature cases, the electron mobility is consistently higher in the 1D simulations. Notably, the disparity in the values of $\mu$ is particularly significant in the cases of positive curvature. This is a noteworthy observation considering that the 1D simulations do not resolve the azimuthal instabilities and their induced electrons' transport. Even though the Q2D simulations capture the additional contribution of the instabilities, the resulting electrons' mobility in the Q2D simulations is still lower compared to that observed from the 1D simulations.

In the 1D simulations, the electrons' mobility shows a very small variation across the domain and for various curvature cases. In the negative-curvature cases, the mobility exhibits a slight increase near the walls, whereas, in the positive-curvature cases, it becomes slightly more pronounced toward the center. Nevertheless, in the Q2D simulations, the electrons' mobility has a notable non-uniform distribution across the domain, and its values vary significantly with the field's curvature. The mobility is particularly significant at $\theta$ = -45 and -60 degrees.

For the case of xenon propellant, we assess, as an example, the importance of various contributions to the axial electron transport. To this end, we consider the individual force terms in the electrons' momentum equation along the azimuthal direction ($z$)

$$-qn_e v_{e,y} B_x = \frac{\partial}{\partial t}(mn_e v_{e,z}) + \frac{\partial}{\partial x}(mn_e v_{e,x} v_{e,z}) + \frac{\partial}{\partial x}(\Pi_{e,xz}) - q\tilde{n}_e \tilde{E}_z - qn_e v_{e,x} B_y \qquad \text{(Eq. 4)}$$

In Eq.4, $q$ is the elementary charge, $m$ is the electron mass, $n_e$ is the electron number density, $v_{e,x}$ and $v_{e,z}$ are the electrons' radial and azimuthal drift velocities, and $\tilde{n}_e$ and $\tilde{E}_z$ are the number density and the azimuthal electric field fluctuations. In this equation, the left-hand-side term is the radial magnetic force term ($F_{Bx}$). On the right-hand side, the first and the second terms represent the temporal inertia ($F_t$) and the convective inertia ($F_I$) force terms, respectively. The third force term ($F_\Pi$) corresponds to the viscous effects, the fourth term ($F_E$) represents the contribution of the azimuthal instabilities, and the last term is the axial magnetic force term ($F_{By}$).

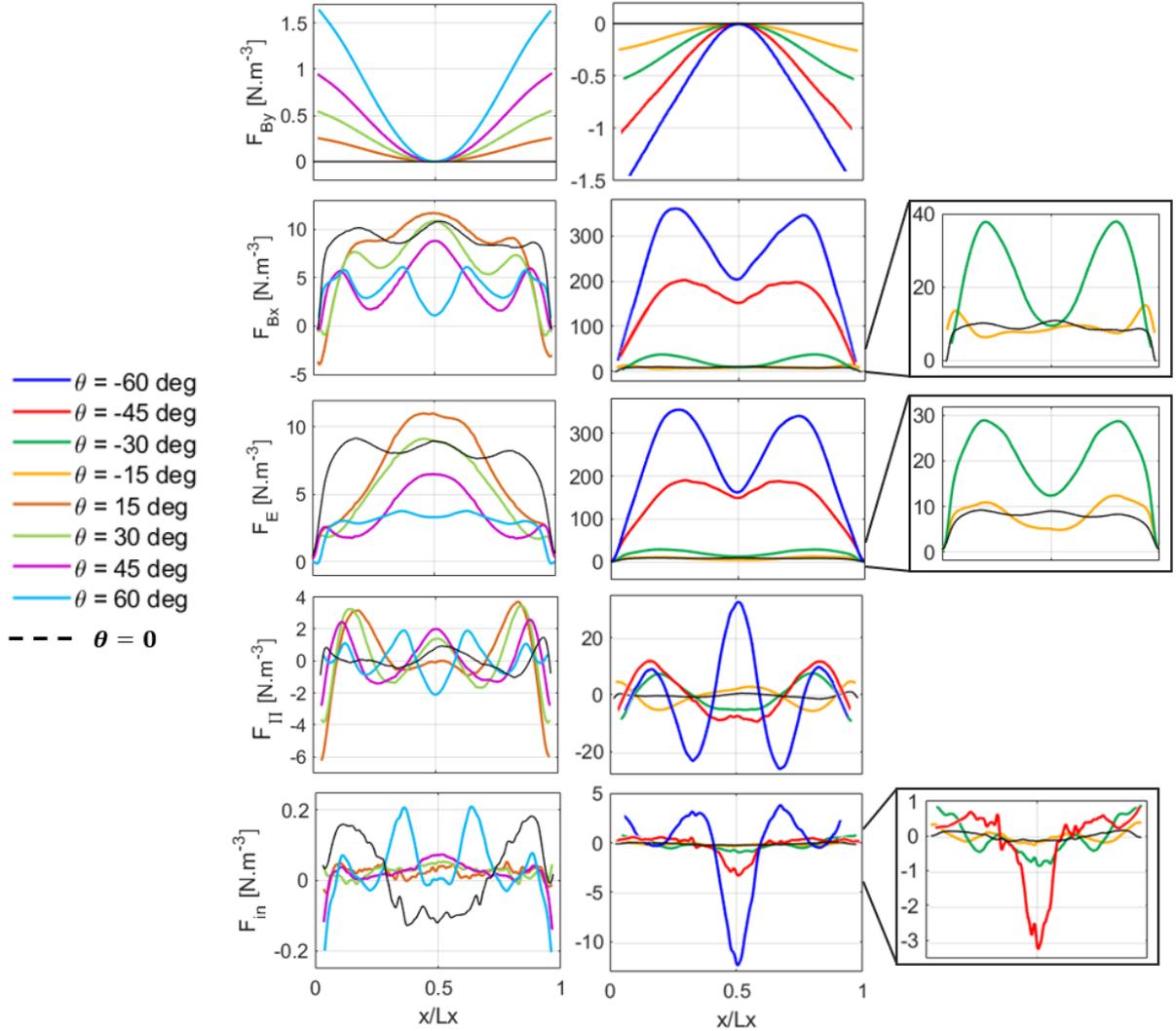

Figure 20: Radial distribution of each of the force terms in Eq. 4 for the xenon propellant and various curvatures of the magnetic field. The momentum terms are averaged over 10 $\mu s$ of the simulations' time (20-30 $\mu s$). Left-column plots are for the positive curvatures, and the right-column plots for the negative curvatures.

The time-averaged radial distributions of each of the force terms in Eq. 4 for various $\theta$ values are compared in Figure 20. Also, the variation with the field's curvature of the radially averaged terms are plotted in Figure 21.

In these two figures, the temporal inertia ($F_t$) is not shown as it was found to have a negligible influence on the electrons' transport in the simulations. From the plots in Figure 20 and Figure 21, we notice that the instabilities are the major contributor to the axial transport. In this regard, we refer to Refs. [35][36], where we demonstrated in a similar radial-azimuthal configuration to that of this work that the radial profiles of the axial electron current density induced by different instability modes are different. We showed that, in the cases where the ECDI is the dominant instability, the axial electron current density peaks at the centerline, whereas the MTSI's dominance leads to the concentration of the electron current near the walls [35][36].

Consistent with these previous observations, we can see from Figure 20 that, across all negative $\theta$ values where the MTSI was seen to be dominant from the spectral analyses, $F_E$ is larger near the walls and reduces toward the center of the domain. Moreover, at $\theta$ = -45 and -60 degrees, $F_E$ is remarkably higher than the cases with $\theta$ = -15 and -30 degrees. This behavior was also observed in Ref. [26], when the long-$\lambda$ instability mode was seen to be strong within the system.

In the cases with positive $\theta$ values of 15 to 45 degrees, where the ECDI was shown in Section 3.2 to be the dominant instability and the MTSI was absent for xenon, the distribution of $F_E$ is mostly concentrated in the center of the domain. However, in the case of $\theta$ = 60 degrees as well as in the baseline case, the $F_E$ distribution becomes more uniform in line with the coexistence of the ECDI and the MTSI.

Figure 21 shows that the second significant contributor to the electrons' transport across several curvature degrees is the pressure term ($F_\Pi$), which shows an oscillatory behavior across the radius of the domain from Figure 20 for all $\theta$ values. Moreover, we also see from Figure 20 that the axial magnetic field force ($F_{By}$) becomes non-negligible toward the walls for large positive and negative curvatures, which is because, at elevated $\theta$ values, the axial component of the magnetic field becomes notable in the regions adjacent to the walls. Indeed, for $\theta$ = 60 degrees, Figure 21 shows that the radial average of the $F_{By}$ term is even more significant than the pressure term. Finally, the inertia force term ($F_I$) plays the least significant role to the axial electron transport in the most cases.

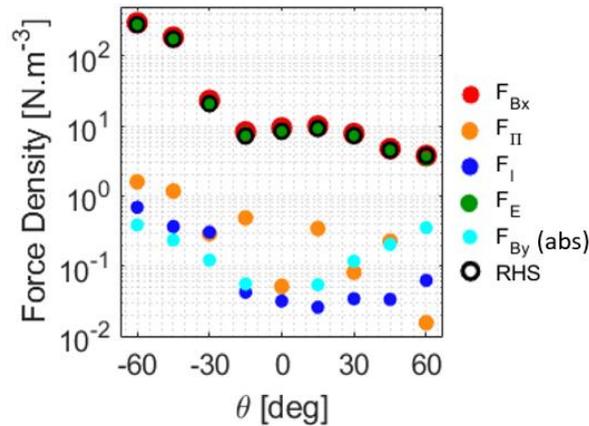

Figure 21: Variation vs $\theta$ of the radially averaged value of each force term in the electrons' azimuthal momentum equation (Eq. 4) from the radial-azimuthal simulations with xenon propellant. The hollow black dot denotes the sum of all terms on the right-hand side of Eq. 4.

### 3.4. Magnetic field's curvature and the ion beam divergence

An interesting impact of the magnetic field's curvature, particularly from an applied perspective, pertains to the ion beam divergence. To quantify the level of divergence from our simulations, we defined an average divergence angle of the ion beam ($\alpha_{avg}$) as the normalized integral of a novel quantity that we term the "divergence-angle distribution function", $f(\alpha, x)$. $\alpha_{avg}$, calculated as per Eq. 5, is a function of the radial coordinate ($x$),

$$\alpha_{avg}(x) = \max\left(-\frac{\int_{-\frac{\pi}{2}}^{0} \alpha f(\alpha,x)d\alpha}{\int_{-\frac{\pi}{2}}^{0} f(\alpha,x)d\alpha}, \frac{\int_{0}^{\frac{\pi}{2}} \alpha f(\alpha,x)d\alpha}{\int_{0}^{\frac{\pi}{2}} f(\alpha,x)d\alpha}\right). \tag{Eq. 5}$$

The parameter $\alpha$ is the angle between the ion particles' trajectory and the axial axis. The ions' "divergence-angle distribution function" or $I\alpha DF$ characterizes the distribution of the ion trajectory angles ($\alpha$) across the radial extent of the domain in the form of a histogram.

According to the ion temperature profiles shown in Figure 8 and the observed values of the $T_i$, especially at large negative curvatures which implied a possible broadening of the ions' velocity distribution function, the radial and azimuthal velocity components of the ions can be expected to contribute to the divergence of the ion beam. Accordingly, we define $\alpha_{rad}$ and $\boldsymbol{\alpha_{azim}}$ as the projection of $\alpha$ on the axial-radial and axial-azimuthal planes, respectively, i.e.,

$$\alpha_{rad} = \tan^{-1}\left(\frac{v_{i,x}}{v_{i,y}}\right), \qquad \alpha_{azim} = \tan^{-1}\left(\frac{v_{i,z}}{v_{i,y}}\right). \tag{Eq. 6}$$

In Eq.6, $v_{i,x}$, $v_{i,y}$, and $v_{i,z}$ are, respectively, the ion particles' velocity along the radial, axial, and azimuthal directions.

We have integrated the ions' radial $I\alpha DF$ ($f_r(\alpha_{rad}, x)$) and their azimuthal $I\alpha DF$ ($f_{az}(\alpha_{azim}, x)$) along the radial direction according to Eq. 7, and the results in terms of $F_r$ vs $\alpha_{rad}$ and $F_{az}$ vs $\alpha_{azim}$ are shown in Figure 22.

$$F_r(\alpha_{rad}) = \int_0^{L_x} f_r(\alpha_{rad}, x)dx, \qquad F_{az}(\alpha_{azim}) = \int_0^{L_x} f_{az}(\alpha_{azim}, x)dx. \tag{Eq. 7}$$

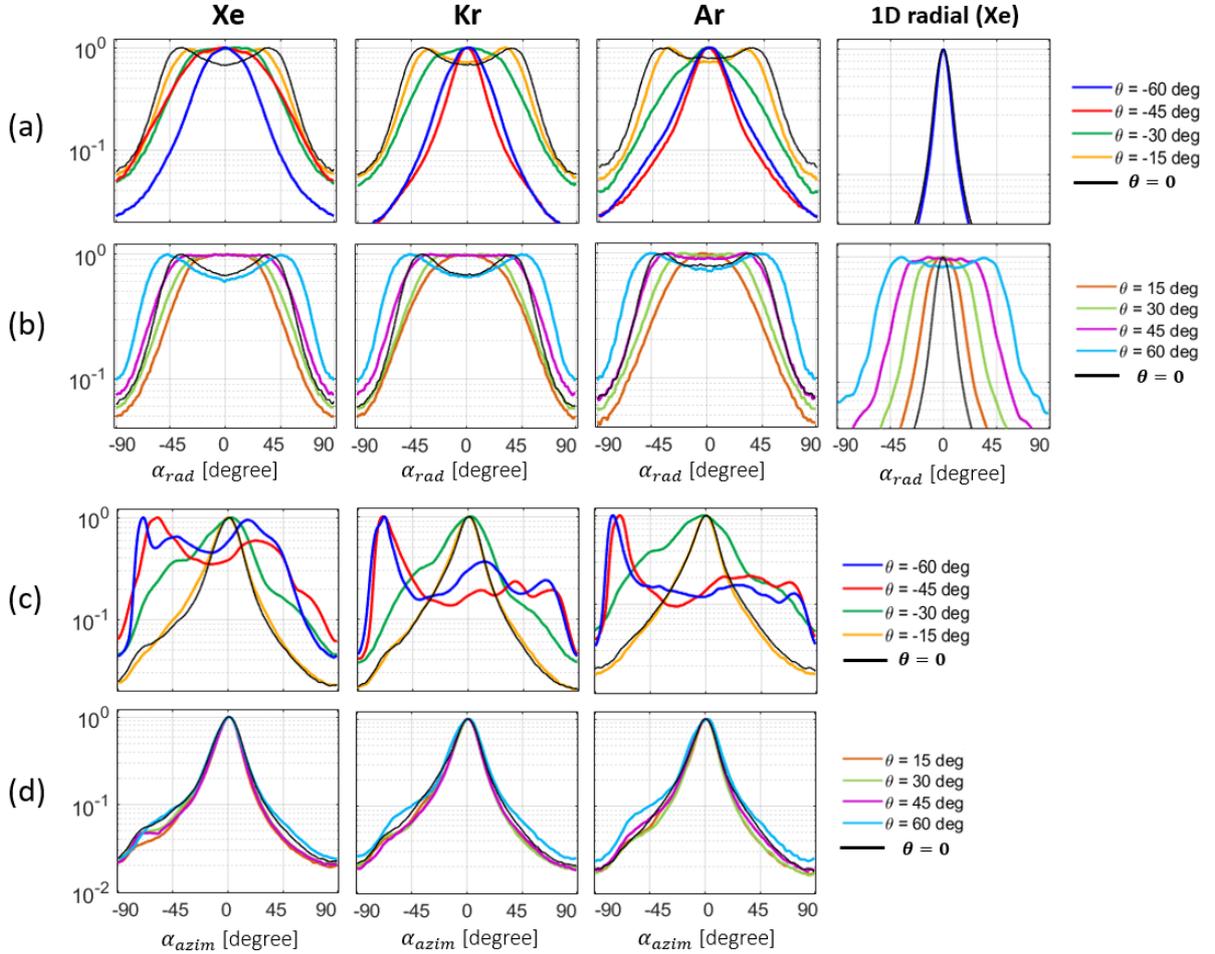

Figure 22: Radially averaged divergence-angle distribution functions for various magnetic field curvatures due to the radial ($F_r$, first and second rows) and azimuthal ($F_{az}$, third and fourth rows) velocity components of the ions for the quasi-2D simulations with various propellants, and due to the radial velocity component of the ions for the 1D radial simulations with Xe; (a) and (c) negative magnetic field curvatures, (b) and (d) positive magnetic field curvatures.

For the Q2D simulations, we see from the first- and second-row plots of Figure 22 that, across all curvature cases, the $F_r$ demonstrates a spread over a broader range of angles compared to the 1D simulations. For the negative-curvature cases, in addition, as $\theta$ varies from 0 to -60 degrees, apart from the decrease in the dispersion of $F_r$ across the velocity angles, the two peaks in the $F_r$ around $\alpha_{rad} = \pm(30\text{-}45)$ degrees merge and form one peak at $\alpha_{rad} = 0$. For the positive curvatures (second-row plots in Figure 22), as $\theta$ increases from 15 to 45 degrees, the peak of the $F_r$ at about $\alpha_{rad} = 0$ start to further flatten, extending over a wider range of radial velocity angles. At $\theta = 60$ degrees, the $F_r$ peaks at about $\alpha_{rad} = \pm 45$ degrees.

Regarding the radially averaged azimuthal $I\alpha DF$ ($F_{az}$), we observe from the third-row plots in Figure 22 that the negative curvature of the field increases the population of the ions with non-zero azimuthal velocity angles. In fact, at $\theta = -45$ and $-60$ degrees, there is a notable ion population having the velocity angle of about $\alpha_{azim} = -55$ to $-70$ degrees. In contrast, the impact of the positive curvature on the $F_{az}$ is not substantial as noticed from the fourth-row plots in Figure 22.

Now, we use our definition of the average divergence angle ($\alpha_{avg}$), given by Eq.5, to obtain the radial distribution of the average ion beam divergence due to radial and azimuthal ions' velocity components. The resulting plots for various $B$-field curvatures and the different propellants from the Q2D simulations are presented in Figure 23. In addition, in this figure, the profiles of the average radial ion divergence angle from the 1D simulations with Xe are provided.

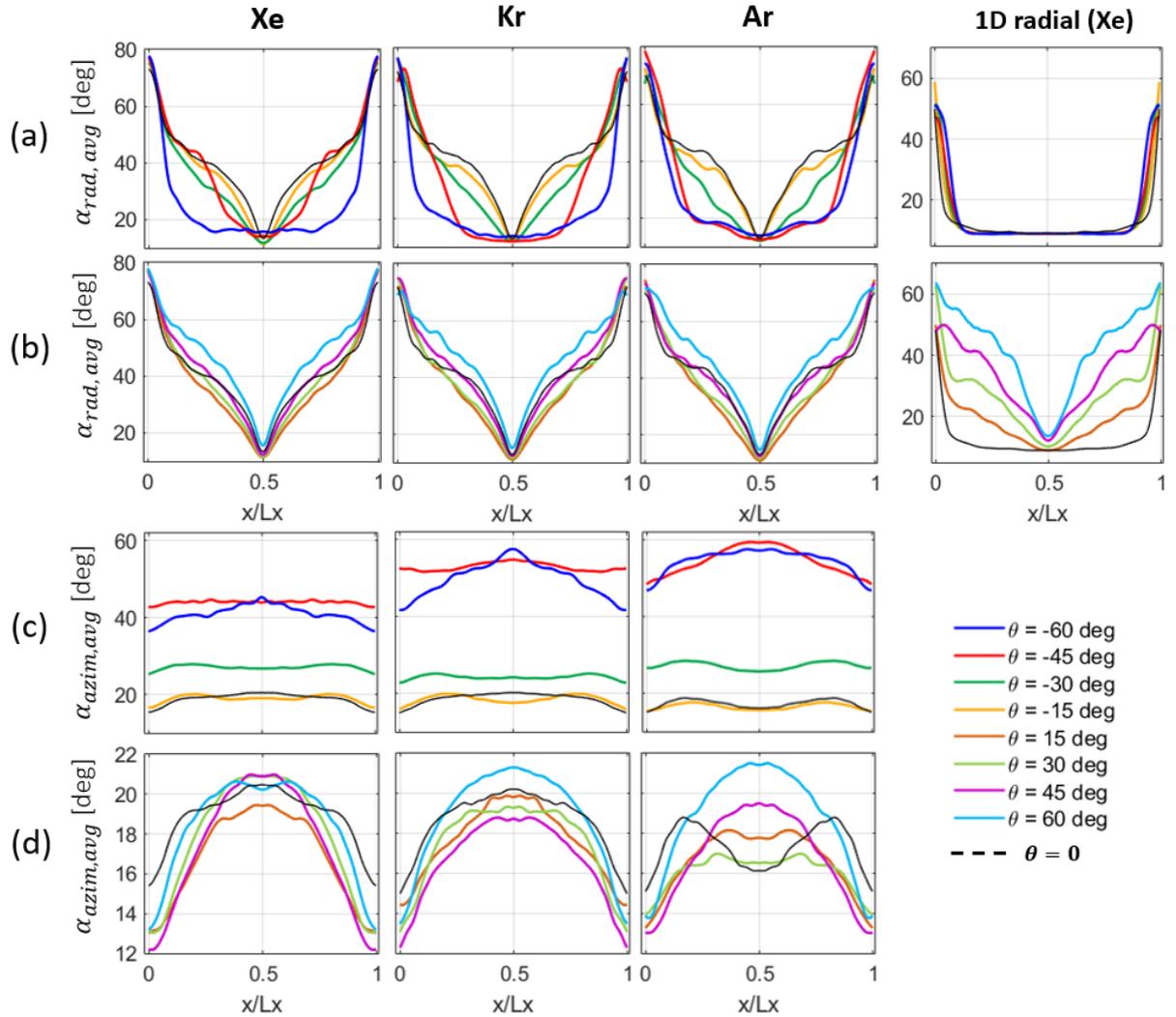

Figure 23: Radial distribution of the average ion beam divergence angle ($\alpha_{avg}$, Eq. 5) for various magnetic field curvatures from the quasi-2D simulations with various propellants and from the 1D radial simulations with Xe; (a) divergence angle due to the ions' radial velocity component ($\alpha_{rad}$) in simulations with negative magnetic field curvatures, (b) divergence angle due to the ions' radial velocity component ($\alpha_{rad}$) in simulations with positive magnetic field curvatures, (c) divergence angle ($\alpha_{azim}$) due to the ions' azimuthal velocity component in simulations with negative magnetic field curvatures, (d) divergence angle due to the ions' azimuthal velocity component ($\alpha_{azim}$) in simulations with positive magnetic field curvatures.

Starting with the ion divergence angle caused by their radial velocity ($\alpha_{rad,avg}$), the plots for the Q2D simulations (Figure 23(a)) show that, in general, the negative curvature of the magnetic field tends to reduce the ion divergence $\alpha_{rad,avg}$ in the central part of the domain. This effect becomes consistently more significant toward higher negative curvatures. It is also observed that the case with $\theta$ = -45 degrees displays a notable disparity in the $\alpha_{rad,avg}$ profiles among different propellants. Specifically, the central ion beam, which can be noticed as a low radial divergence angle, expands over a larger portion of the domain's radial extent going from Xe to Ar. In the 1D simulations, however, the ion beam radial divergence does not exhibit significant sensitivity to the degree of negative $B$-field curvature. Also, contrary to the Q2D simulations, in the 1D setting, the divergence angle remains low in the plasma bulk away from sheaths.

Looking at the $\alpha_{rad,avg}$ plots for the positive curvature cases (Figure 23(b)), we notice that the positive curvature expectedly enhances the ions' radial divergence angle. This effect is notably stronger in the 1D simulations than in the Q2D simulations.

Referring to Figure 23(c) and (d), we assess the ion beam divergence due to the azimuthal component of the ions' velocity ($\alpha_{azim,avg}$) from the Q2D simulations. It is observed that, overall, the negative curvature significantly increases $\alpha_{azim,avg}$, which, for the negative $\theta$s, assumes a relatively uniform distribution across the radial extent of the domain. Furthermore, in the cases of $\theta$ = -60 and -45 degrees, there is a consistent rise in $\alpha_{azim,avg}$ from Xe to Ar. On the contrary, as depicted in Figure 23(d), the positive curvature of the magnetic field moderately impacts the radial profile of $\alpha_{azim,avg}$. Notably, the relationship between $\alpha_{azim,avg}$ and the field's curvature displays a non-monotonic trend and exhibits distinct behaviors across various propellants.

To summarize the observed trends, considering the mean of the radial and azimuthal divergence angles ($\alpha_{rad,mean}$ and $\alpha_{azim,avg}$) over the entire domain's radius as representative values to quantify the total ion beam divergence, we plotted the variations against $\theta$ of the $\alpha_{rad,mean}$ and $\alpha_{azim,avg}$ for the Q2D simulations in Figure 24. The mean radial divergence angle ($\alpha_{rad,mean}$) from the 1D simulations is also plotted against $\theta$ in Figure 24.

Overall, for the Q2D simulations with high negative field's curvature ($\theta$ = -60 and -45 degrees), the beam divergence is primarily driven by the azimuthal ion velocity. This highlights another significance of resolving the azimuthal direction in the Hall thruster simulations, namely, that in the absence of the azimuthal physics, there is a risk of underestimating the ion beam divergence in the situations of a concave field topology with large degrees of curvature.

In any case, across the moderate negative curvatures as well as the entire positive curvature range, the radial ion velocity has the dominant influence on the ion beam divergence. Moreover, in the 1D simulations, the ion divergence remains almost invariant across the negative field curvatures, whereas it demonstrates an almost linear increase over the positive curvature values.

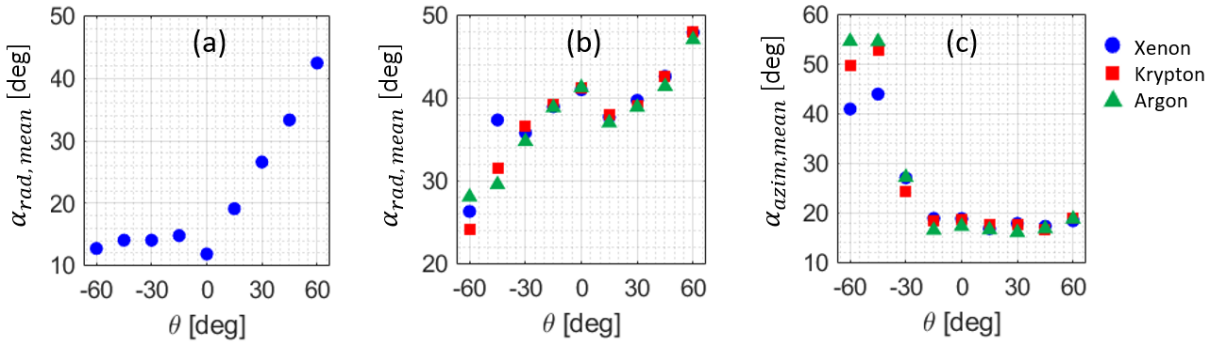

Figure 24: Variation vs $\theta$ of the mean divergence angle; (a) divergence angle due to the ions' radial velocity component in 1D simulations, (b) divergence angle due to the ions' radial velocity component in Q2D simulations. (c) divergence angle due to the ions' azimuthal velocity component in Q2D simulations.

**Section 4: Conclusions**

The variation in the radial-azimuthal dynamics of the plasma in a Hall thruster due to various degrees of negative and positive magnetic field curvature was investigated in detail in this article for three propellants, xenon, krypton, and argon. We presented the results of reduced-order quasi-2D simulations for each propellant at multiple curvature degrees and compared the main observations against the results of the 1D radial simulations with xenon.

We assessed the effects of the magnetic field's curvature on the spatial distribution of the time-averaged plasma properties, the instability spectra, the electrons' mobility characteristics, and the divergence of the ion beam due to the radial and azimuthal velocity components that can be affected by the underlying instability mechanisms. For Xe, we also compared the contributions of various momentum force terms to the cross-field transport of the electrons.

In terms of the most salient findings, we first observed that, comparing the 1D and Q2D simulations, the overall predicted behaviors of the plasma properties for the positive-curvature cases were rather similar. Contrarily, for the negative-curvature cases, there had been specific notable differences between the 1D and the Q2D simulations in terms of the plasma properties' distribution. In addition, the effects of the field's curvature in the 1D simulations were often more exaggerated.

Second, the peak ion number density within the domain was seen to reduce with the increasing degree of the field's curvature from $\theta$ = -60 to 60 degrees. In both the Q2D and the 1D simulations, the peak $n_i$ values dropped significantly across the negative curvatures, whereas it showed a very moderate variation with the positive curvatures. This trend was consistently observed for all propellants. The radial electron temperature ($T_{ex}$) from the Q2D simulations showed an increase in values from $\theta$ = -60 to -30, followed by a monotonic decrease with $\theta$ from -30 to 45 degrees and then a slight increase at $\theta$ = 60 degrees. The 1D simulations' trend was, however, similar to the $n_i$ variations with $\theta$ as was observed from the 1D cases. The ion temperature from the Q2D simulations exhibited large values of 35-45 eV for highly negative field's curvature ($\theta$ = -60 and -45 degrees), consistent with the observations reported in Refs. [26][29]. The $T_i$'s radial profile was distinctly different in shape between the positive and the negative values of the curvature in both the 1D and the Q2D simulations.

Third, the radial position of the ion sonic point was noticed, from the Q2D simulations, to continuously shift away from the vicinity of the lateral walls as $\theta$ increased from -60 to 60 degrees. The significance of this observation was highlighted with regard to determining the correct location of the sheath's edge in the simulations that involve curved magnetic field topologies.

Fourth, the electron temperature anisotropy in terms of the radial-to-azimuthal temperature ratio ($T_{ex}/T_{ez}$) from the Q2D simulations demonstrated a distinct and insightful variation with the field's curvature, which was generally shared among all propellants. Overall, this ratio assumed values larger than 1 for the negative curvatures and values less 1 for the zero- and positive-curvature cases up to $\theta$ = 45 degrees. These observations were correlated to the dominance of various instability modes within the discharge.

Following on the above, and from the detailed spectral analyses of the instabilities that we conducted using the FFT and the DMD methods, we confirmed that, in general, the concavity or convexity of the magnetic field curvature can lead to the establishment of two distinct plasma regimes in a Hall thruster and, by extension, in the E×B plasma configurations. These regimes are characterized by the relative dominance of either the MTSI or the ECDI modes. Accordingly, the negative curvature of the field was seen to suppress the ECDI and the purely azimuthal instabilities except for the azimuthal long-$\lambda$ mode, which was observed across the negative-curvature cases. In addition, the negative-curvature condition was noted to promote the development of dominant instabilities in the system similar to the MTSI with wavenumbers along both the radial and the azimuthal directions. The presence of a positive curvature of the magnetic field was observed to hinder the formation of the MTSI and the instabilities with the radial-azimuthal structures in general. Under these circumstances, the modes were seen to primarily consist of purely azimuthal instabilities, such as the ECDI, with different frequencies and wavenumbers. The consistency between these findings and the variations trend of the average electron temperature anisotropy across the simulated $B$-field curvatures highlighted the suitability of the $T_{ex}/T_{ez}$ ratio to serve as a general criterion for assessing the comparative significance of the ECDI and the MTSI modes.

Another important point to emphasize here is that the DMD analyses presented in this work provide an effective route for the cross-validation of our insights against the experiments. In fact, having an accessible visual representation of the spatial structure (radial and/or azimuthal wavelengths) of various instability modes can inform the design of experimental setups and the involved diagnostics in order to verify the numerical observations reported concerning the variation in the instability spectra vs the degree of the magnetic field's curvature.

Regarding the electrons' mobility, the change in the degree of the field's curvature was noticed to affect the radial distribution and the magnitude of this property in line with the variations in the instability spectra and the dominance of different modes. The results were almost identical for different propellants. In terms of the values and the radial profiles, the mobility from the Q2D simulations showed a much higher degree of sensitivity to the $B$-field's curvature compared to the 1D simulations, especially for the negative curvatures. For xenon and from the Q2D simulations, the evaluation of the individual role in cross-field transport of various force terms in the electrons' azimuthal momentum equation underlined the significant contribution of the instabilities through the electric force term. The viscous force term as well as the axial magnetic force term were also seen to play a non-negligible role in certain curvature conditions and/or spatial locations over the radial extent of the domain.

Concerning the divergence of the ion beam with respect to the axial direction, we pursued a novel approach to quantify the divergence induced by the radial and azimuthal components of the ions' velocity. Most notably, we observed from the Q2D simulations that, even though a concave field topology reduces the ion beam's divergence along the radial direction, at highly negative curvature angles, the divergence is dominantly determined by the azimuthal velocity component and is significant in values due to the influence of the azimuthal instabilities. At the moderate negative field's curvatures and, for the positive curvatures, the ion beam divergence is governed by the radial velocity component and expectedly increases with the curvature degree.

The insights provided in this work from multiple facets into the effects of the magnetic field's curvature, in addition to the notably different physical processes and behaviors observed for the cases with the negative and the positive curvatures, show, by the way of conjecture, that, apart from the role of the axial gradients in the $B$-field, the curvature of the magnetic field can be an additional important reason behind the differences observed in Hall thrusters' plasma dynamics before and after the $B$-field's peak. The validity of this speculation is of course contingent on further verification of our results in a 3D PIC simulation setup that resolves the axial physics as well as the ionization process self-consistently.

**Acknowledgments**:

The present research is carried out within the framework of the project "Advanced Space Propulsion for Innovative Realization of space Exploration (ASPIRE)". ASPIRE has received funding from the European Union's Horizon 2020 Research and Innovation Programme under the Grant Agreement No. 101004366. The views expressed herein can in no way be taken as to reflect an official opinion of the Commission of the European Union.

The authors gratefully acknowledge the computational resources and support provided by the Imperial College Research Computing Service (http://doi.org/10.14469/hpc/2232).

**Data Availability Statement**:

The simulation data that support the findings of this study are available from the corresponding author upon reasonable request.